\documentclass[a4paper,11pt]{article}
\pdfoutput=1 

\usepackage{jcappub} 

\usepackage[T1]{fontenc} 
\usepackage{lineno}

\title{Cosmic-muon characterization and annual modulation measurement with Double Chooz detectors}

\collaborationImg{\includegraphics[width=.15\textwidth]{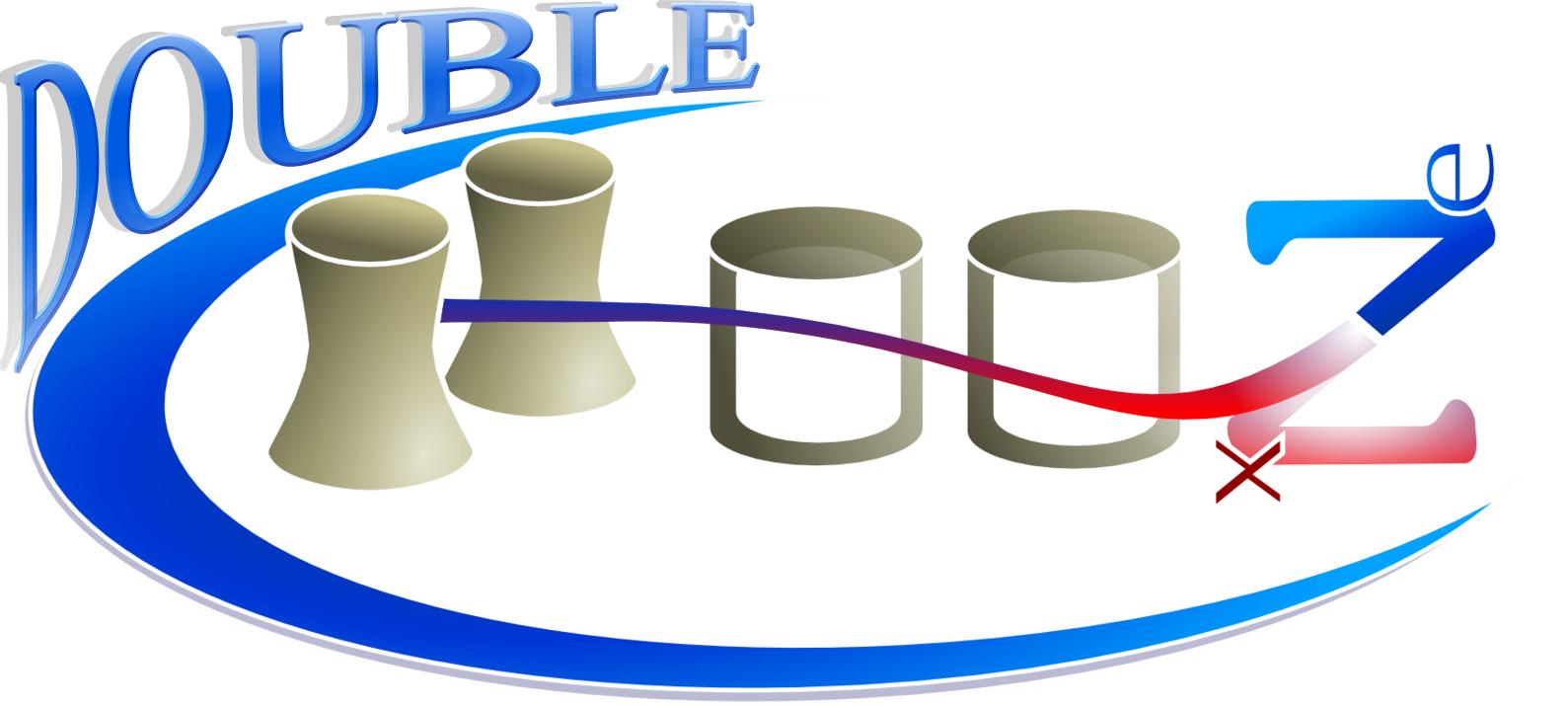}}
\author{The Double Chooz collaboration \\}
\author[a,2]{T. Abrah\~ao,}  
\author[b]{H. Almazan,}  
\author[a]{J.C. dos Anjos,} 
\author[c]{S. Appel,} 
\author[d]{E. Baussan,} 
\author[e]{I. Bekman,} 
\author[f,g]{T.J.C. Bezerra,} 
\author[h]{L. Bezrukov,} 
\author[i]{E. Blucher,} 
% ...

\affiliation[a]{Centro Brasileiro de Pesquisas F\'isicas, Rio de Janeiro, RJ, 22290-180, Brazil}
\affiliation[b]{Max-Planck-Institut f\"ur Kernphysik, 69117 Heidelberg, Germany}
\affiliation[c]{Physik Department, Technische Universit\"at M\"unchen, 85748 Garching, Germany}
\affiliation[d]{IPHC, Universit\'e de Strasbourg, CNRS/IN2P3, 67037 Strasbourg, France}
\affiliation[e]{III. Physikalisches Institut, RWTH Aachen University, 52056 Aachen, Germany}
\affiliation[f]{SUBATECH, CNRS/IN2P3, Universit\'e de Nantes, Ecole des Mines de Nantes, 44307 Nantes,France}
\affiliation[g]{Research Center for Neutrino Science, Tohoku University, Sendai 980-8578, Japan}
\affiliation[h]{Institute of Nuclear Research of the Russian Academy of Sciences, Moscow 117312, Russia}
\affiliation[i]{The Enrico Fermi Institute, The University of Chicago, Chicago, Illinois 60637, USA}
% ...

\emailAdd{hgomez@apc.univ-paris7.fr}

\abstract
{
A study on cosmic muons has been performed for the two identical near and far neutrino detectors of the Double Chooz experiment, placed at $\sim$120 and $\sim$300 m.w.e. underground respectively, including the corresponding simulations using the MUSIC simulation package. This characterization has allowed us to measure the muon flux reaching both detectors to be  (3.64 $\pm$ 0.04) $\times$ 10$^{-4}$ cm$^{-2}$s$^{-1}$ for the near detector and (7.00 $\pm$ 0.05) $\times$ 10$^{-5}$ cm$^{-2}$s$^{-1}$ for the far one. The seasonal modulation of the signal has also been studied observing a positive correlation with the atmospheric temperature, leading to an effective temperature coefficient of $\alpha_{T}$ = 0.212 $\pm$ 0.024 and 0.355 $\pm$ 0.019 for the near and far detectors respectively. These measurements, in good agreement with expectations based on theoretical models, represent one of the first measurements of this coefficient in shallow depth installations.
}


\arxivnumber{1611.07845}

\begin{document}
\maketitle
\flushbottom

\section{Introduction}
\label{sec:Introduction}

Muons created by cosmic rays in the atmosphere and muon-induced events are one of the main concerns regarding background in neutrino experiments. The impact of these events on the research of the expected signals is usually reduced by placing the experimental set-ups in deep underground facilities. However, a precise identification of these events for further rejection is required, especially in cases where the detector is installed on the surface or at shallow depth, with the muon flux remaining high. Total expected rate, flux and muon angular distributions are some of the information that can be valuable for muon characterization. Empirically, some of this information can already be obtained by various methods and by the development of dedicated muon reconstruction algorithms optimized for the detector features. This information can be complemented by performing corresponding simulations.

\begin{figure}[tbp]
  \centering 
  \includegraphics[width=.45\textwidth]{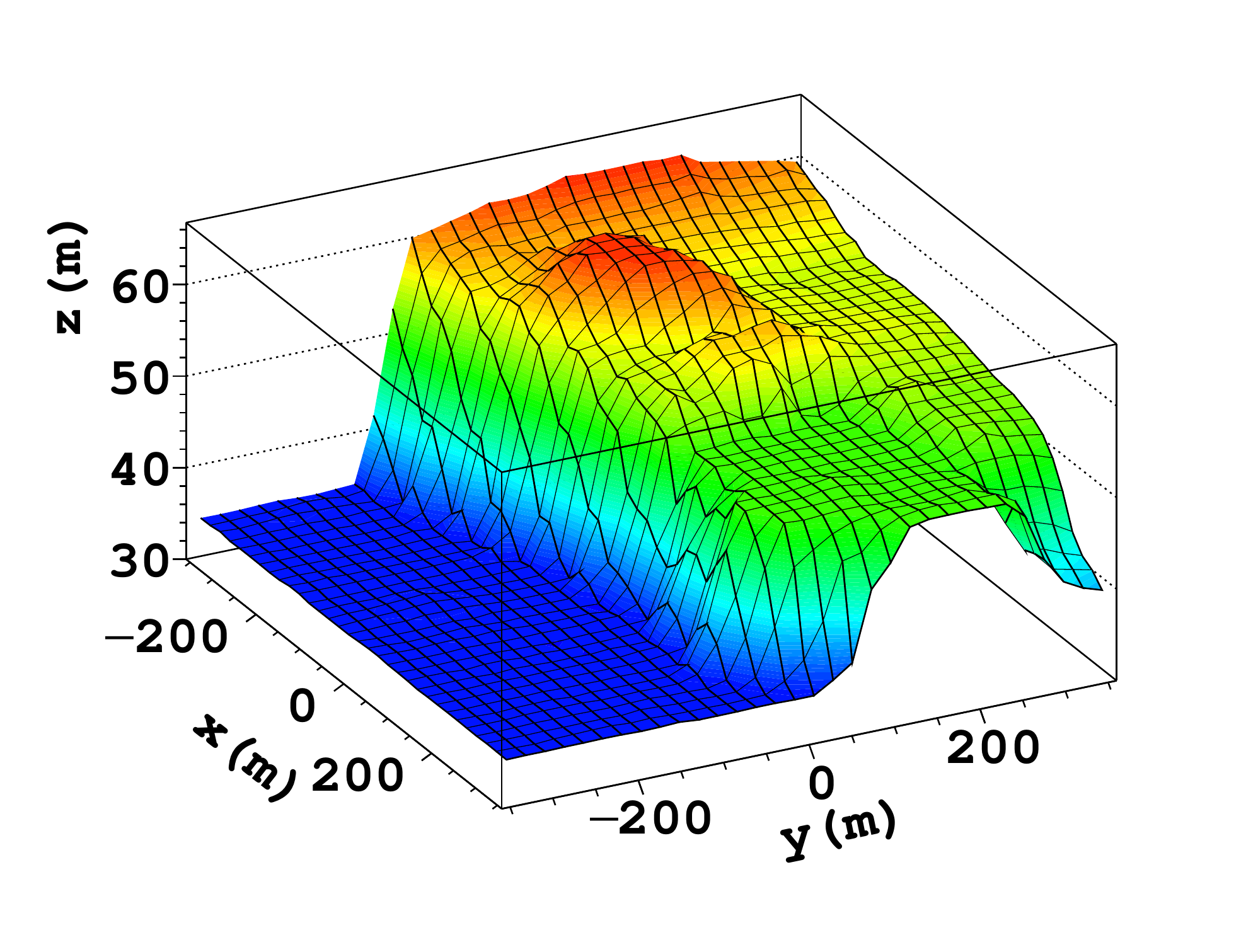}
  \includegraphics[width=.45\textwidth]{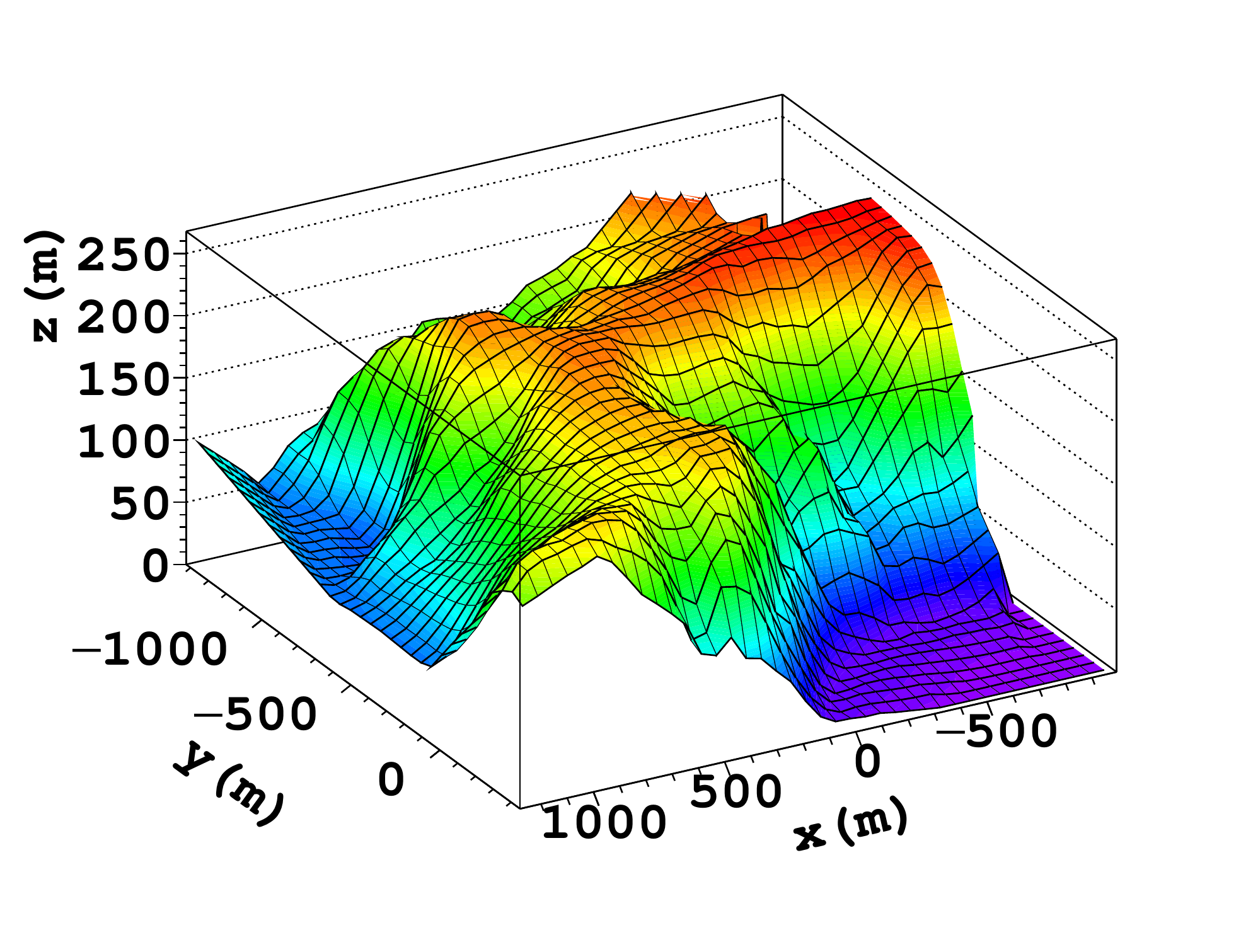}
  \caption{Digitized profile of the Double Chooz near (left) and far (right) detectors' overburden. The detector centre is located at the coordinate origin.}
  \label{Profiles}
\end{figure}

From the digitized overburdens of Double Chooz near and far detectors showed in figure \ref{Profiles} and their composition, the corresponding mean depths can be estimated as $\sim$120 and $\sim$300 m.w.e. respectively. These low overburdens imply that a significant quantity of muons still reaches the detectors. In fact, most of the background events recorded by Double Chooz detectors are directly related to muons \cite{DChooz}. Furthermore, for these shallow depth installations, the overburden profile should also be taken into account. Profile irregularities can imply that the corresponding depth for a given direction would be significantly different from the mean value. In the case of muons this effect implies two main consequences: inhomogeneities in the muon incident angle distribution, and differences in the energy threshold of muons reaching the detector with respect to the estimated value from the mean overburden. These effects are more identifiable in the case that the overburden decreases significantly for a given direction (as it is the case of the far detector as observed in Figure \ref{Profiles}), since for this direction more muons would be detected and the corresponding muon energy threshold will be lower than the estimated by the mean depth, implying that lower energy muons should be taken into account in the corresponding muon studies.

Because both Double Chooz experimental halls are shallow depth installations, an accurate characterization of muons reaching the corresponding detectors is important, considering the above mentioned particularities. Taking advantage of the long period of time that Double Chooz detectors have been taking data ($\sim$151 days for the near detector and $\sim$673 days for the far one), it is possible to perform, for the first time, an extended study of muons at shallow depth installations, including the estimation of the muon flux and the corresponding angular distributions. Moreover, the long data taking period also allows the study of the seasonal effect on the muon rate. Due to the temperature variations in the atmosphere during the year, the density of the atmosphere is expected to vary, being lower in summer and increasing the mesons' mean free path. This effect leads to an expected increase of muons reaching the detectors in summer, when the temperature is higher. The correlation between detected muon rate and atmospheric temperature permits the estimation of the effective temperature coefficient ($\alpha_{T}$). It can be compared with theoretical estimations for a lower muon energy threshold (22.3 GeV for the near detector and 46.0 GeV for the far one based on simulations described in section \ref{sec:MuonSimulations}) than those studied by experiments located deep underground, with energy thresholds closer to the TeV scale. These studies can be complemented by corresponding simulations, whose comparison with experimental data provides a validation of the muon transport simulation in a wider energy range, since lower energy muons should be taken into account in Double Chooz.

With all these objectives, diverse muon studies based on the experimental data from both detectors and on simulations have been conducted. After presenting the main features of the Double Chooz detectors (section \ref{sec:DoubleChooz}), results from muon analyses are presented in section \ref{sec:MuonFluxAngularDistributions}, while the corresponding simulations and the comparison with the experimental results are reported in section \ref{sec:MuonSimulations}. Finally, the analysis focused on the modulation of the muon signal and the atmospheric effective temperature, as well as their correlation, is presented in section \ref{sec:AnnualModulation} before summarizing the main results (section \ref{sec:Conclusions}).

\section{Double Chooz detectors}
\label{sec:DoubleChooz}

\begin{figure}[tbp]
\centering 
\includegraphics[width=.5\textwidth]{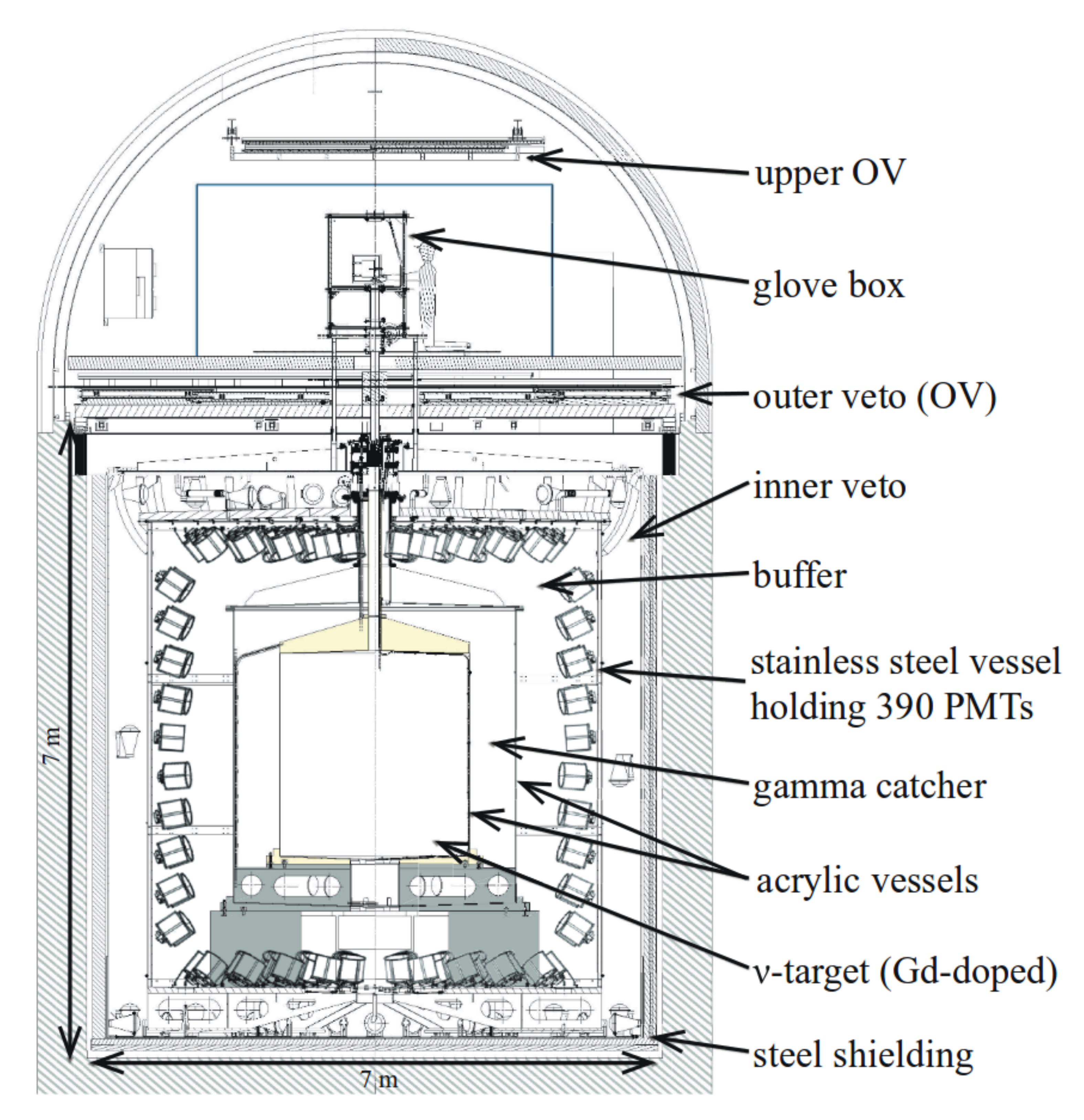}
\caption{Double Chooz detectors layout. From inner to outer: neutrino target (NT), gamma catcher (GC), buffer tank and inner veto (IV).}
\label{fig:DCLayout}
\end{figure}

Double Chooz operates two equivalent detectors, placed at $\sim$400 and $\sim$1050 m from the Chooz nuclear power plant reactors respectively, to study the oscillation of the electron anti-neutrinos ($\bar{\nu}_{e}$) generated in the reactor cores. These detectors are composed of a set of concentric cylinders with an outer muon veto on top. The neutrino target (NT) is the innermost volume and contains 8.3 tons of Gd-loaded (0.1 \%) liquid scintillator inside a transparent acrylic vessel, where the neutrinos interact via the inverse beta-decay (IBD) process. The NT is surrounded by another acrylic vessel, the gamma catcher (GC), filled with unloaded scintillator, which is in turn surrounded by a buffer tank filled with mineral oil. The GC allows the full containment of the energy deposition of gamma rays from the neutron capture on Gd and the positron annihilation in the NT region. The buffer volume acts as a shield against the radioactivity of the detector surroundings, including that coming from the 390 photomultiplier tubes (PMTs) installed for light collection. The NT, GC and the buffer tank form the inner detector (ID), which is surrounded by the inner veto (IV), a 50 cm thick liquid scintillator volume equipped with 78 PMTs. The ID and IV are in turn surrounded by 1 m of water in the near detector and 15 cm of steel in the far one. Finally, the upper part of the detector is covered by an outer muon veto (OV), consisting of plastic scintillator strips grouped into modules. Figure \ref{fig:DCLayout} shows the layout of the Double Chooz detectors. A more detailed description can be found in refs. \cite{DChooz, DChooz_LN} and references therein.

\section{Muon flux and angular distributions}
\label{sec:MuonFluxAngularDistributions}

\subsection{Muon event selection}

Muons deposit large amounts of energy when they transverse the regions of the detectors filled with liquid scintillator, referred as scintillation volumes. This category encompasses the active volumes of the ID, comprising the NT and GC as explained in section \ref{sec:DoubleChooz}, and the IV. Therefore, the muon selection in Double Chooz relies on the energy deposition in the ID and IV volumes. Due to the Double Chooz detector design, with its concentric cylindrical volumes, two kinds of muon events can be registered: muons depositing energy only in the IV or in both the ID and IV. For the first case muons can be selected applying a sole cut on the measured energy in the IV (E$_{\textrm{vis}}^{\textrm{IV}}$), while for the second case, muons are tagged by the combined cuts on the measured energy in the ID and the IV (E$_{\textrm{vis}}^{\textrm{ID}}$ and E$_{\textrm{vis}}^{\textrm{IV}}$ respectively).

For muons crossing only the IV the applied cut is:

\begin{itemize}
\item E$_{\textrm{vis}}^{\textrm{IV}}$ $>$ 25 MeV
\end{itemize}

while for muons crossing the ID and IV volumes, the deposited energy in both volumes must satisfy the following thresholds:

\begin{itemize}
\item E$_{\textrm{vis}}^{\textrm{ID}}$ $>$ 30 MeV and, 
\item E$_{\textrm{vis}}^{\textrm{IV}}$ $>$ 5 MeV 
\end{itemize}  

By applying these requirements most of the non-muon events, corresponding to low energy deposits, are rejected while the loss of efficiency is almost negligible. If these cut values are increased, especially the E$_{\textrm{vis}}^{\textrm{IV}}$ for muons crossing the IV and E$_{\textrm{vis}}^{\textrm{ID}}$ for muon crossing the ID and IV, a purer muon sample can be obtained, although with a consequent loss of efficiency. However, this can be useful to perform some studies regarding the muon track angular distributions and the muon flux modulation as described in following sections.

Based on this muon selection, and from the analysis of $\sim$151 days of data for the near detector and $\sim$673 days for the far one,  the detected mean rate in Double Chooz detectors of muons crossing at least one of the scintillator volumes has been estimated as 242.75 $\pm$ 4.81 $s^{-1}$ for the near detector and 46.16 $\pm$ 1.04 $s^{-1}$ for the far one.

\subsection{Muon track reconstruction}
\label{MuonTrackReconstruction}

Tracks of all the selected muons are reconstructed by using data from the ID, the IV and the OV, performing a fit using mainly the time information from the ID and IV, and the spatial information from the OV if the  muon has crossed this detector. The algorithm assumes that muons are through-going, but if the energy deposit suggest that a muon may have stopped in the ID, the algorithm also performs a fit for this hypothesis and chooses between the two via the relative goodness-of-fit. This algorithm has demonstrated a spatial resolution of $\sim$40 mm for muons crossing the central volume. A detailed description of the algorithm and its performance can be found in ref. \cite{FIDO}.

\begin{figure}[tbp]
  \centering 
  \includegraphics[width=0.80\textwidth]{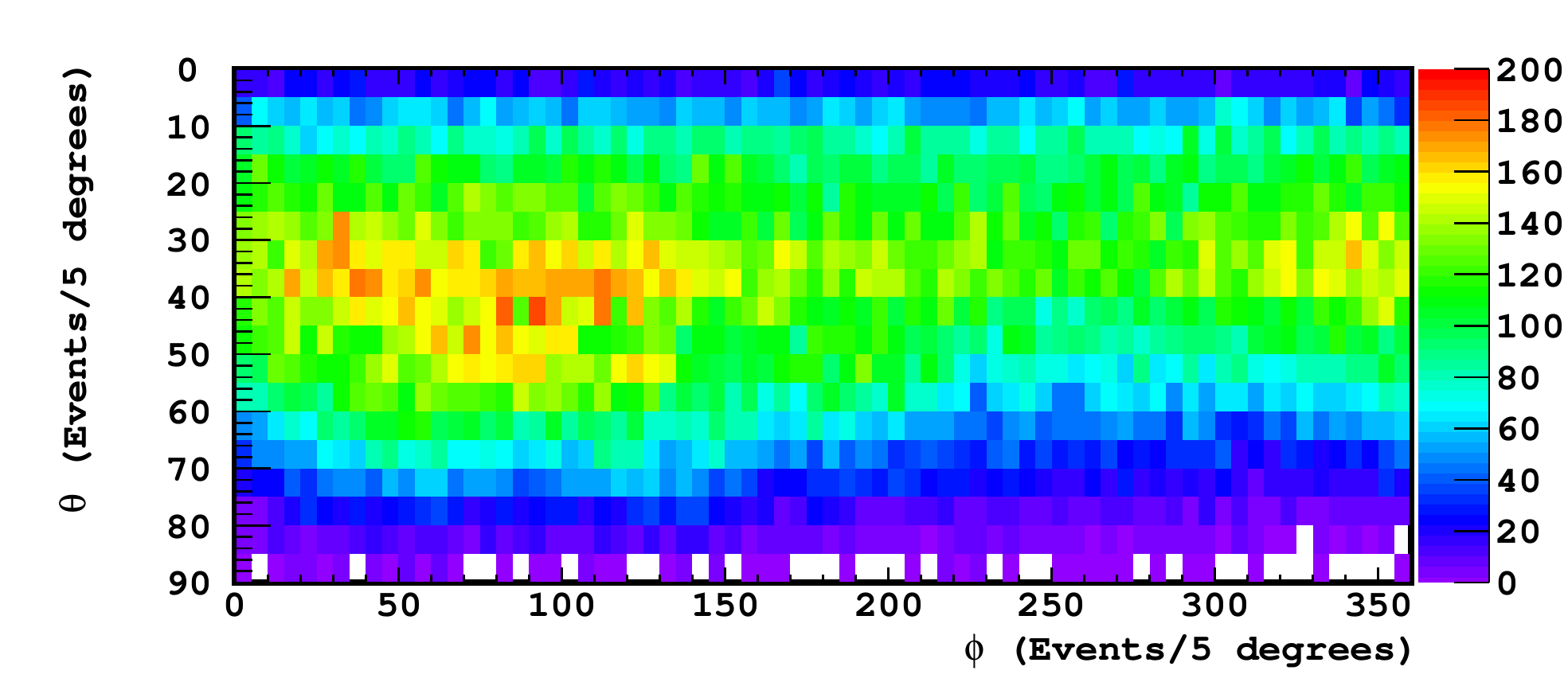}
  \includegraphics[width=0.80\textwidth]{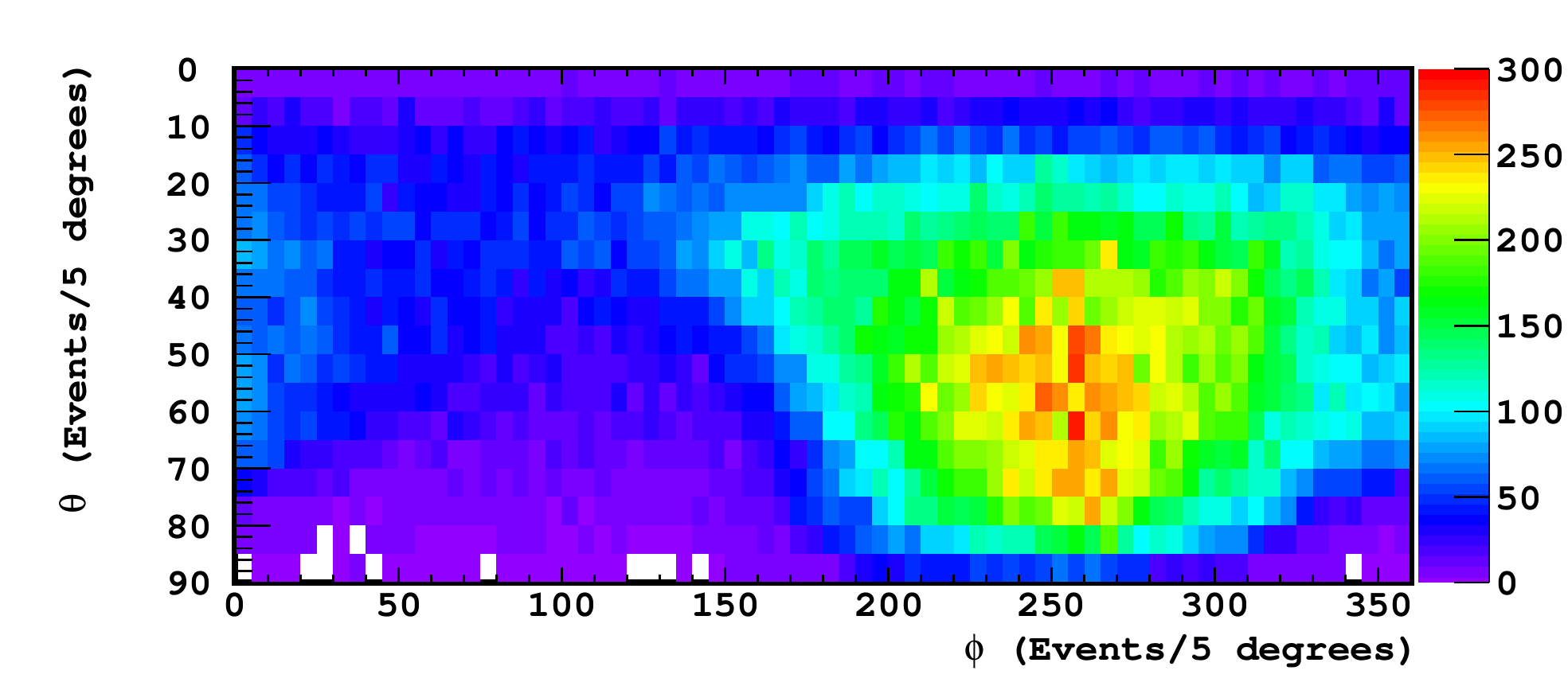}
  \caption{$\phi$-$\theta$ angular distributions corresponding to 10$^5$ reconstructed muon tracks from experimental data reaching the Double Chooz near (top) and far (bottom) detectors.}
  \label{Muon_data_output_ThetaPhi}
\end{figure}

In order to ensure that the reconstructed tracks correspond to muons having a substantial path length within the detector, for which the reconstruction algorithm has greater precision, the muon selection has been restricted by increasing the threshold of the deposited energy in the ID to E$_{\textrm{vis}}^{\textrm{ID}}$ $>$ 200 MeV. Figure \ref{Muon_data_output_ThetaPhi} presents the reconstructed direction of muons reaching the detectors depending on the zenith ($\theta$) and azimuth ($\phi$) angles. For both detectors the spherical coordinate system is centered at the detector centre, with the azimuthal direction $\phi$ = 0$^{\circ}$ pointing to the experimental hall entrance. These distributions behave as expected since more muons are detected for the directions in which the overburden is shallower as described in section \ref{sec:Introduction}. This effect is more pronounced for the far detector since its overburden (figure \ref{Profiles}) is more irregular.

\subsection{Muon flux}

In general, the muon flux is calculated as the ratio between the detected muon rate and the detector effective area. When detectors have spherical geometry, the estimation of the effective area is simpler than for cylindrical detectors, as in the case of Double Chooz. However, tracks of muons crossing the ID are available for the data taking periods. Using this information, the calculation of detector effective area for the muon flux can be simplified.

From the track information, it is possible to estimate the muon track distance of closest approach to the detector centre. From muons crossing the ID (selected by means of the deposited energy), it is possible to select those crossing the detector at a distance to the centre smaller than R (defined as radial cut). By this selection, the detector can be approximated as spherical, with an effective surface area of $\pi R^2$. As mentioned, tracks are reconstructed for those muons crossing the ID and, more specifically, crossing the volumes filled with liquid scintillator (NT and GC), which are contained within the GC cylinder, of 3.4 m diameter and 3.6 m height. With these dimensions, the shortest distance of the GC surface to the detector centre is 1.7 m, and the longest 2.5 m. Therefore, spheres defined having radii between these two values fit better within the GC cylinder.

\begin{figure}[tbp]
  \centering 
  \includegraphics[width=.45\textwidth]{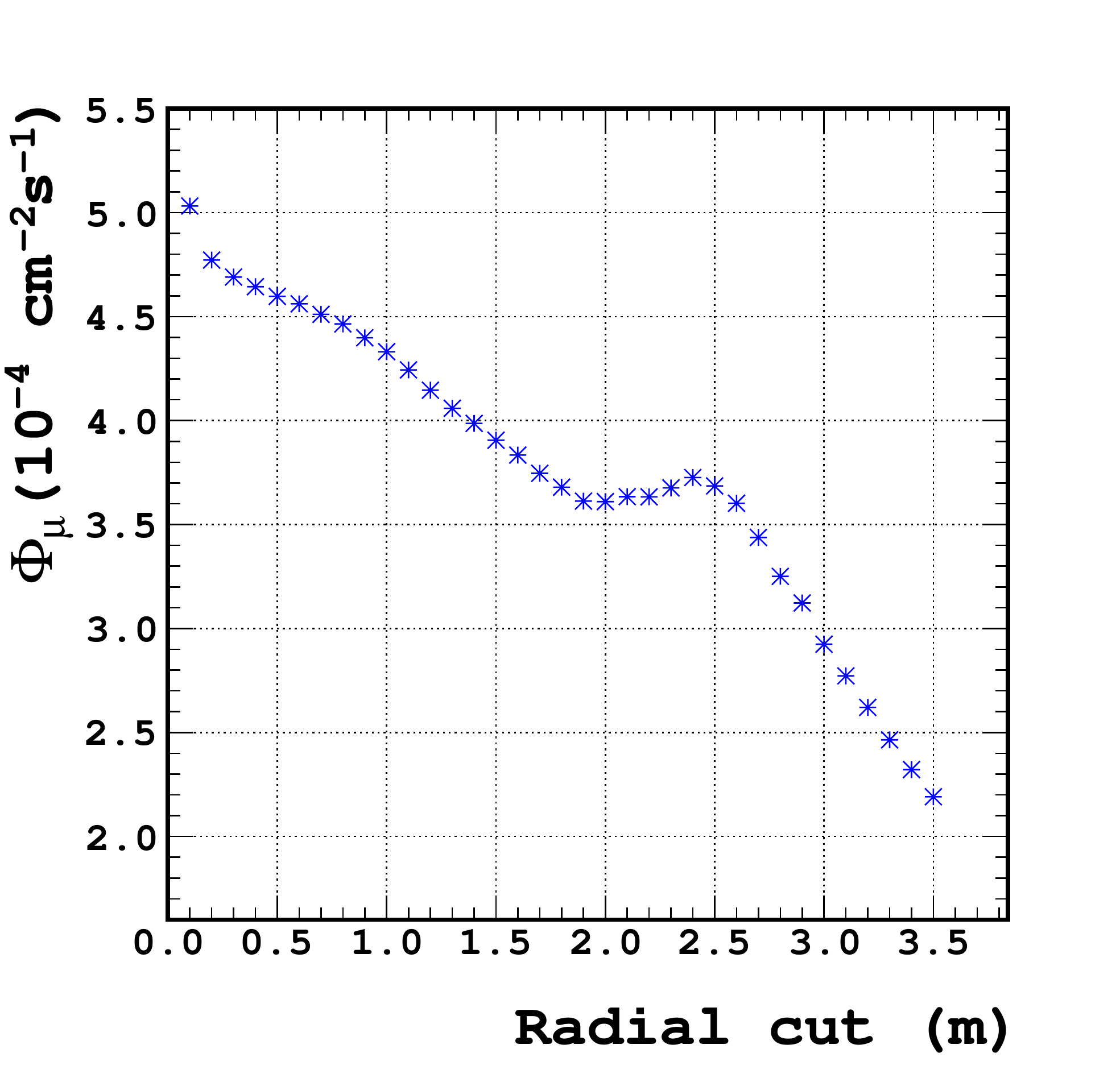}
  \includegraphics[width=.45\textwidth]{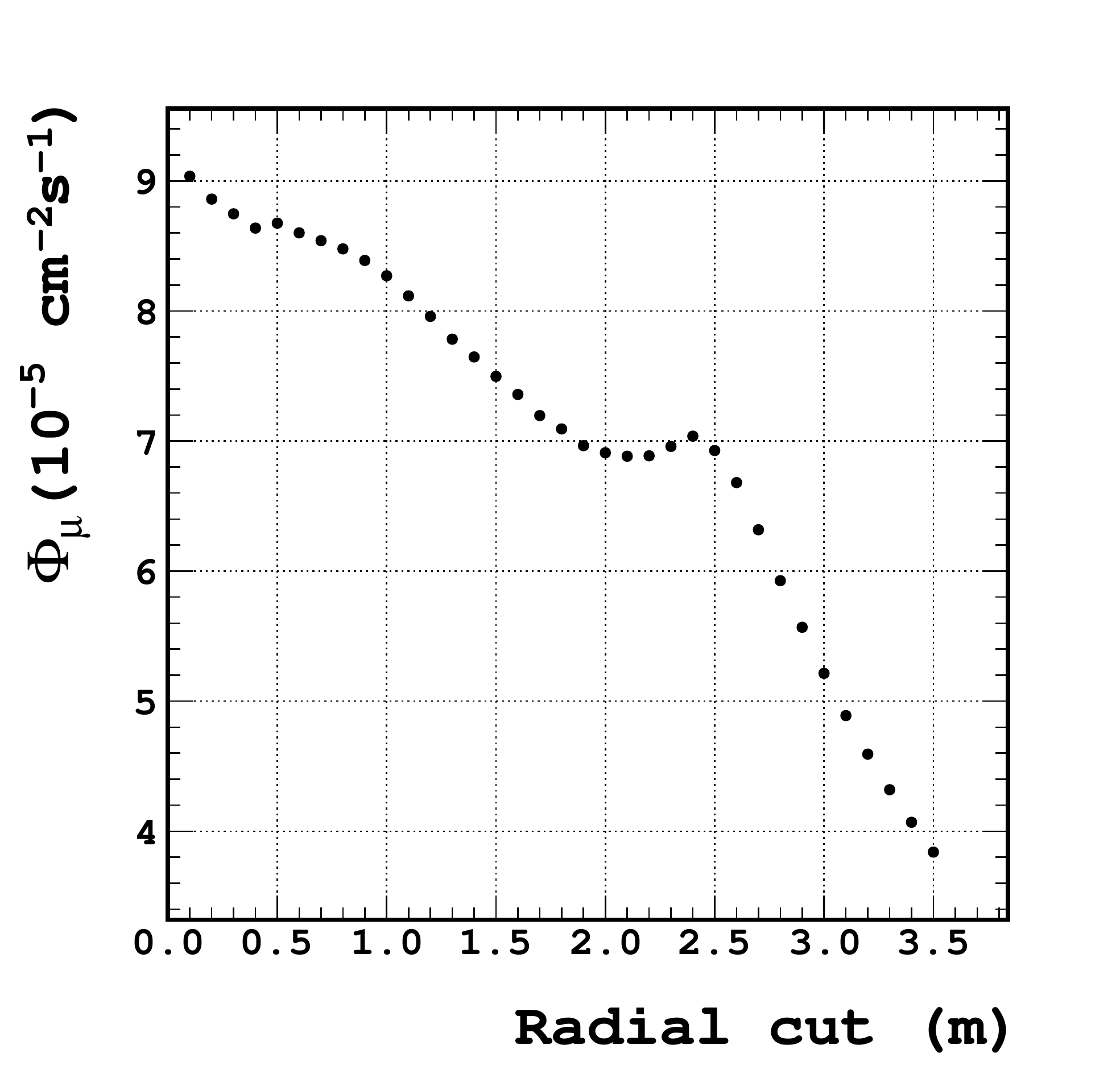}
  \caption{Muon flux as a function of the cut on the muon track distance to the detector centre (radial cut) for the Double Chooz near (left) and far (right) detectors.}
  \label{MuFlux_Sphere}
\end{figure}

Figure \ref{MuFlux_Sphere} shows the reconstructed muon flux as a function of the defined radial cut, revealing an equivalent behaviour for both detectors. The spatial resolution of the reconstruction algorithm induces a systematic inwards radial shift, leading to overestimated flux values when the radial cut value is small. However, for larger radial values, this shift is compensated, yielding a more reliable value for the muon flux. For the radial cut values corresponding to spheres fitting better within the GC (the above mentioned 1.7 and 2.5 m) the reconstructed flux remains constant, indicating that this is the unbiased value. If the radial cut value is bigger than 2.5 m, the flux is underestimated if no efficiency correction is applied as the corresponding sphere no longer fits within the GC limits. Based on these features, the muon flux has been estimated as the mean value of the reconstructed flux in the radial cut window between 1.7 and 2.5 m. It results in $\Phi_{\mu}$=(3.64 $\pm$ 0.04) $\times$ 10$^{-4}$ cm$^{-2}$s$^{-1}$  for the near detector and $\Phi_{\mu}$=(7.00 $\pm$ 0.05) $\times$ 10$^{-5}$ cm$^{-2}$s$^{-1}$ for the far one. The large amount of  analyzed data (more than 10$^9$ selected muons available for both detectors) makes the statistical error negligible. The systematic uncertainties have two contributions. First one, associated with the detection efficiency, is estimated as 0.07\% for both detectors. Additionally, the precision on the distance to the centre distribution induces a systematic which has been estimated by varying the radial cut window. This leads to an additional 1.1\% systematic for the near detector and 0.7\% for the far one. Both contributions lead to overall systematic uncertainty in the muon flux estimation of 1.1\% and 0.7\% for the near and far detectors respectively.

\section{Muon simulations}
\label{sec:MuonSimulations}

The muon flux reaching the detectors and the corresponding angular distributions can also be estimated by performing Monte Carlo simulations. Other information that cannot be obtained from data analysis, such as the muon energy, can also be extracted from them. With this aim, a set of simulations to characterize muons reaching the Double Chooz detectors has been performed. To do this, the MUSIC simulation package has been used \cite{MUSIC}. It is a FORTRAN code for the simulation of the 3D transport of muons through matter taking into account the energy loss due to ionization, pair production, bremsstrahlung and inelastic scattering. MUSIC is capable of simulating the muon energy loss, angular deviation, and lateral displacement. MUSIC requires two main inputs to carry out the simulation: the description of the overburden profile and its composition, and the initial muon distribution at the surface (muon energy and incident direction). 

\subsection{Overburden profile and composition}
\label{OverburdenProfileAndComposition}

The profiles of both experimental halls' overburden have been parametrized using the detector centre as coordinate origin as already showed in figure \ref{Profiles}. Both profiles have been digitized over an area large enough to enable study of muons with incident zenith angles up to at least 75 degrees, reaching 90 degrees in those azimuthal directions where the overburden is shallower, especially for the far detector. This ensures the simulation of practically all the muons capable of reaching the detectors. The main difference in the overburden digitizations of the two detectors comes from the density of points. For the far detector the overburden has been generated from a 2D contour map using the shareware code 3DField \cite{3DField}. Moreover, in the regions where the profile is more irregular, more measurements have been performed and digitized in order to obtain a more accurate definition of the profile. For the near detector, the overburden has been defined from a topographic plan provided by the \textit{Institut national de l'information g\'{e}ographique et foresti\`{e}re} (IGN \cite{IGN}). The plan provides a  less precise mesh than for the far detector. This definition can induce some inaccuracies in the simulation that should be taken into account.

Regarding the material composition, for the far detector, a rock type based on the chemical composition of the Ardennes rock, where the detector is located, has been defined. For the near detector another type of rock has been used based on geological studies performed during the construction of the near detector experimental hall and its access. Table \ref{DC_Materials} summarizes the main properties of both materials as defined.

\begin{table}[tbp]
\centering
\begin{tabular}{|l|cc|}
\hline
& Near detector & Far detector \\ 
\hline
$\langle Z \rangle$ & 11 & 11.8\\
$\langle A \rangle$ & 22 & 24.1\\
$\rho$ (g cm$^{-3}$) & 2.65 & 2.81 \\
$\lambda$ (g cm$^{-2}$) & 26.5 & 23.3 \\
\hline
\end{tabular}
\caption{Summary of the main properties of the overburden material for both detectors: mean atomic number ($\langle Z \rangle$), mean atomic weight ($\langle A \rangle$), density ($\rho$) and radiation length ($\lambda$).}
\label{DC_Materials}
\end{table}

\subsection{Muon distribution at surface}
\label{MuonDistributionAtSurface}

The muon distribution at the surface represents a critical input for the simulation since the result obtained will directly depend on this distribution. For these simulations, an extended version of the Gaisser parametrization \cite{gaisser1990cosmic} has been used. The original parametrization is adequate for muon energies $E_{\mu}$ $\geq$ 100/cos($\theta$) GeV and incident zenith angles $\theta$ $\leq$ 70$^{\circ}$. Due to the depth of the Double Chooz installations and the geometry of the corresponding overburdens, the modifications described in \cite{Tang_PhysRevD_2006} have been taken into account to consider muons with lower energies and higher incident angles.

\begin{figure}[tbp]
  \centering 
  \includegraphics[width=0.8\textwidth]{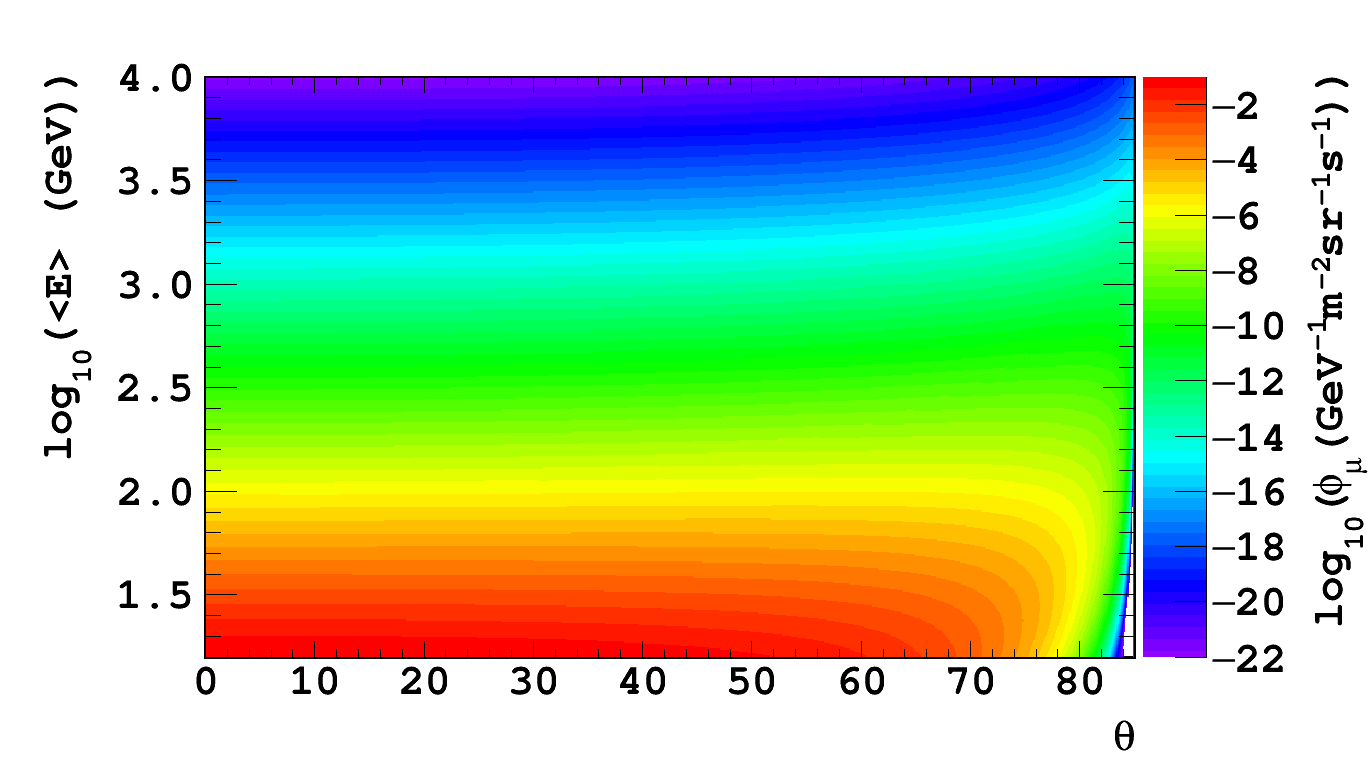}
  \caption{Distribution of the muon energy versus incident zenith angle ($\theta$) at surface for the extended Gaisser parametrization, considering an energy threshold of 20 GeV as the lowest threshold for simulations.}
  \label{Muon_input_Etheta}
\end{figure}

Based on the overburden profiles and compositions presented in section \ref{OverburdenProfileAndComposition}, it is possible to estimate a theoretical energy threshold for a muon to reach the detector. This value is important to optimize the computing time for simulations. In the case of the near detector, this minimum energy corresponds approximately to $\sim$22 GeV, while for the far detector, it is $\sim$46 GeV. Simulation results will provide a more precise value for these muon energy thresholds. In order to safely cover the muon energy range in the simulations, energy thresholds have been set to 20 and 40 GeV for the near and far detectors respectively. These values are well above the threshold for which the considered muon parametrization is valid, so a direct comparison of the initial muon distributions can be done in both cases. Figure \ref{Muon_input_Etheta} shows the $E_{\mu}$-$\theta$ distribution of the generated muons using the extended Gaisser parametrization for the lowest energy threshold considered (corresponding to the near detector case) and the whole zenith angle range.

\subsection{Simulation results: comparison with experimental results}
\label{Simulation_results}

As mentioned, the MUSIC software has been used to transport the muons at surface to the Double Chooz detectors. From these simulations, the three components of the muon momentum at the detector are stored. In addition, the percentage of simulated muons that reach the detector is also computed. It can be used as a normalization factor in order to estimate the mean muon flux from the corresponding flux at surface. In addition to the flux, with this information it is possible to obtain the energy and angular distributions of the muons reaching the detectors.

The expected muon flux reaching the detector can be computed as the product of the simulated muon flux at the surface and the survival probability obtained from these simulations. Due to the constraints coming from the overburden definition, not all the muons from the 2$\pi$ solid-angle hemisphere are simulated, but only those suitable to traverse the overburden. This excludes muons with high $\theta$ angle which should be removed from the initial muon flux. In addition, the ranges of energies of the simulated muons have been defined by the energy threshold determined by the detector overburdens and should also be taken into account in computing the muon flux at surface. The uncertainties associated with this value for the far detector are bigger since the overburden is less regular than for the near detector. The survival probability is directly related to the overburden geometry and composition. For the far detector the digitization of the profile is precise enough that the obtained value for the survival probability has a negligible error. For the near detector case, due to the lower precision on the digitization, an uncertainty is induced. This has been evaluated by performing simulations shifting the profile by  a distance related to the precision of the measurement points provided by the IGN plan. Table \ref{Sim_Flux_Results} summarizes the values obtained on the initial and final fluxes for both detectors taking into account the corresponding uncertainties. The reconstructed fluxes extracted from the experimental data as described in section \ref{sec:MuonFluxAngularDistributions} are also indicated. For both cases, the reconstructed flux from simulations is in agreement with the experimental data within the associated uncertainties.

\begin{table}[tbp]
\centering
\begin{tabular}{|l|cc|}
\hline
& Near detector & Far detector\\ 
\hline
Simulated muon flux at surface (cm$^{-2}$s$^{-1}$) & (1.03 $\pm$ 0.01) $\times$ 10$^{-3}$  &  (3.65 $\pm$ 0.17) $\times$ 10$^{-4}$ \\
Survival probability $(\%)$ & 33.60 $\pm$ 5.86 & 19.82 \\
Muon flux at detector (cm$^{-2}$s$^{-1}$) (Simulations) & (3.47 $\pm$ 0.12) $\times$ 10$^{-4}$ & (7.24 $\pm$ 0.33) $\times$ 10$^{-5}$ \\
\hline
Muon flux at detector (cm$^{-2}$s$^{-1}$) (Data) & (3.64 $\pm$ 0.04) $\times$ 10$^{-4}$ & (7.00 $\pm$ 0.05) $\times$ 10$^{-5}$ \\
\hline
\end{tabular}
\caption{Estimated muon flux at the Double Chooz near and far detectors based on MUSIC simulations together with the reconstructed muon flux values from experimental data.}
\label{Sim_Flux_Results}
\end{table}

\begin{figure}[tbp]
  \centering 
  \includegraphics[width=0.80\textwidth]{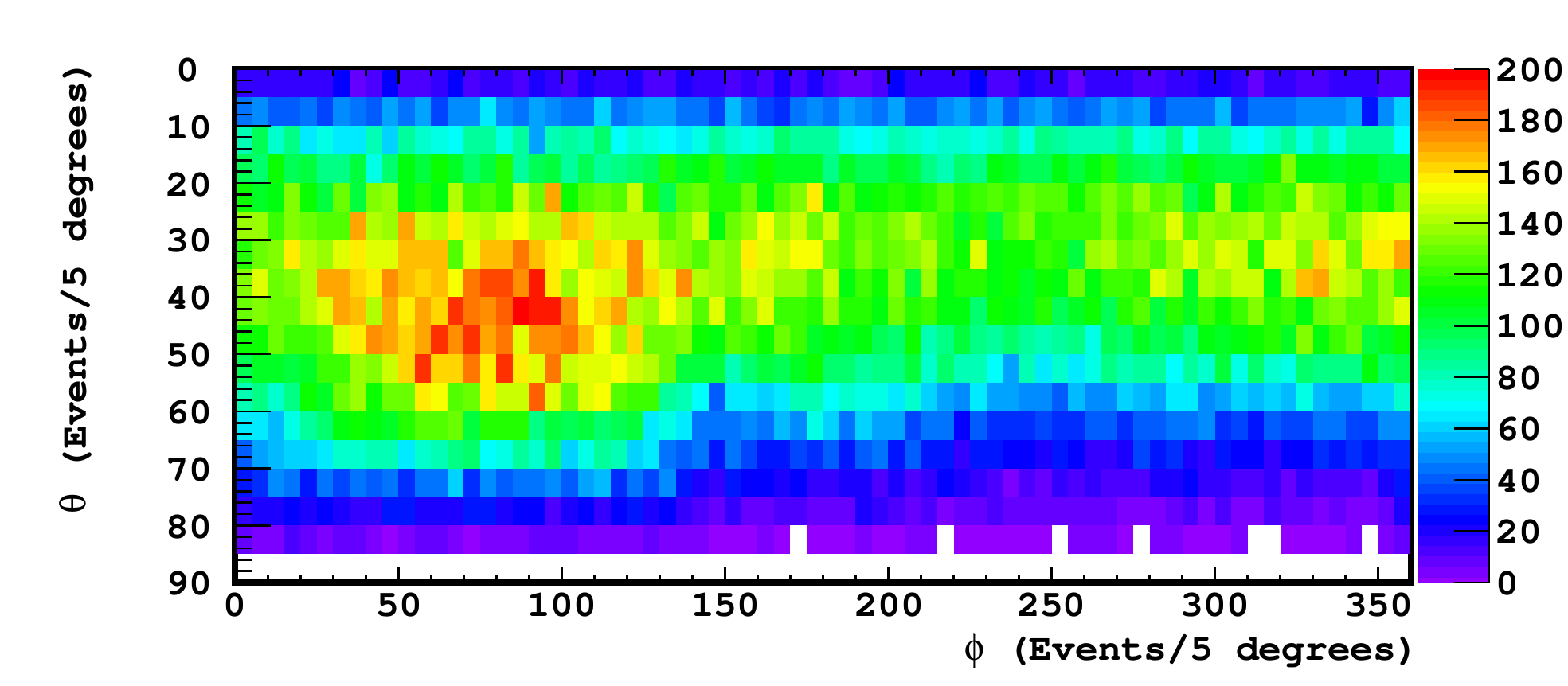}
  \includegraphics[width=0.80\textwidth]{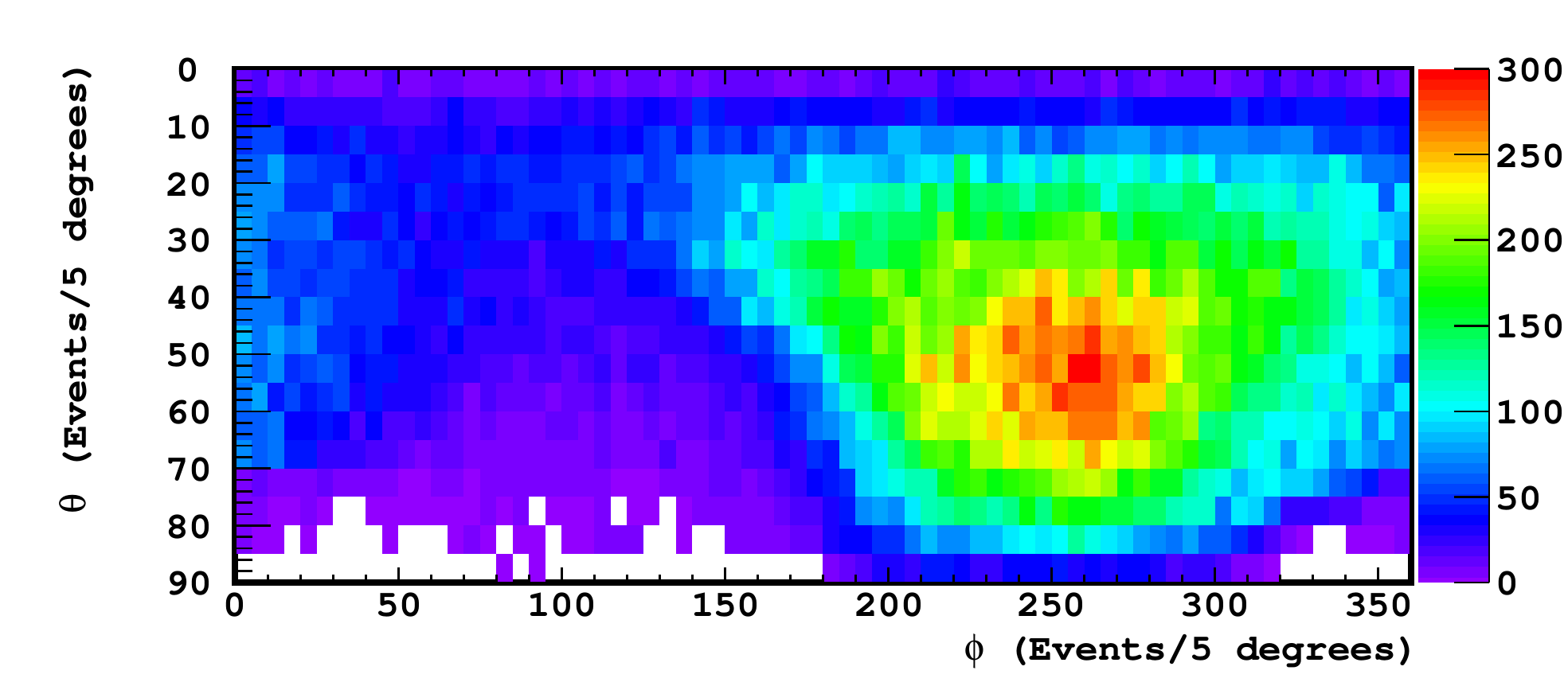}  
  \caption{$\phi$-$\theta$ angular distributions corresponding to 10$^5$ simulated muons reaching the Double Chooz near (top) and far (bottom) detectors.}
  \label{Muon_output_ThetaPhi}
\end{figure}

\begin{figure}[tbp]
  \centering 
  \includegraphics[width=0.45\textwidth]{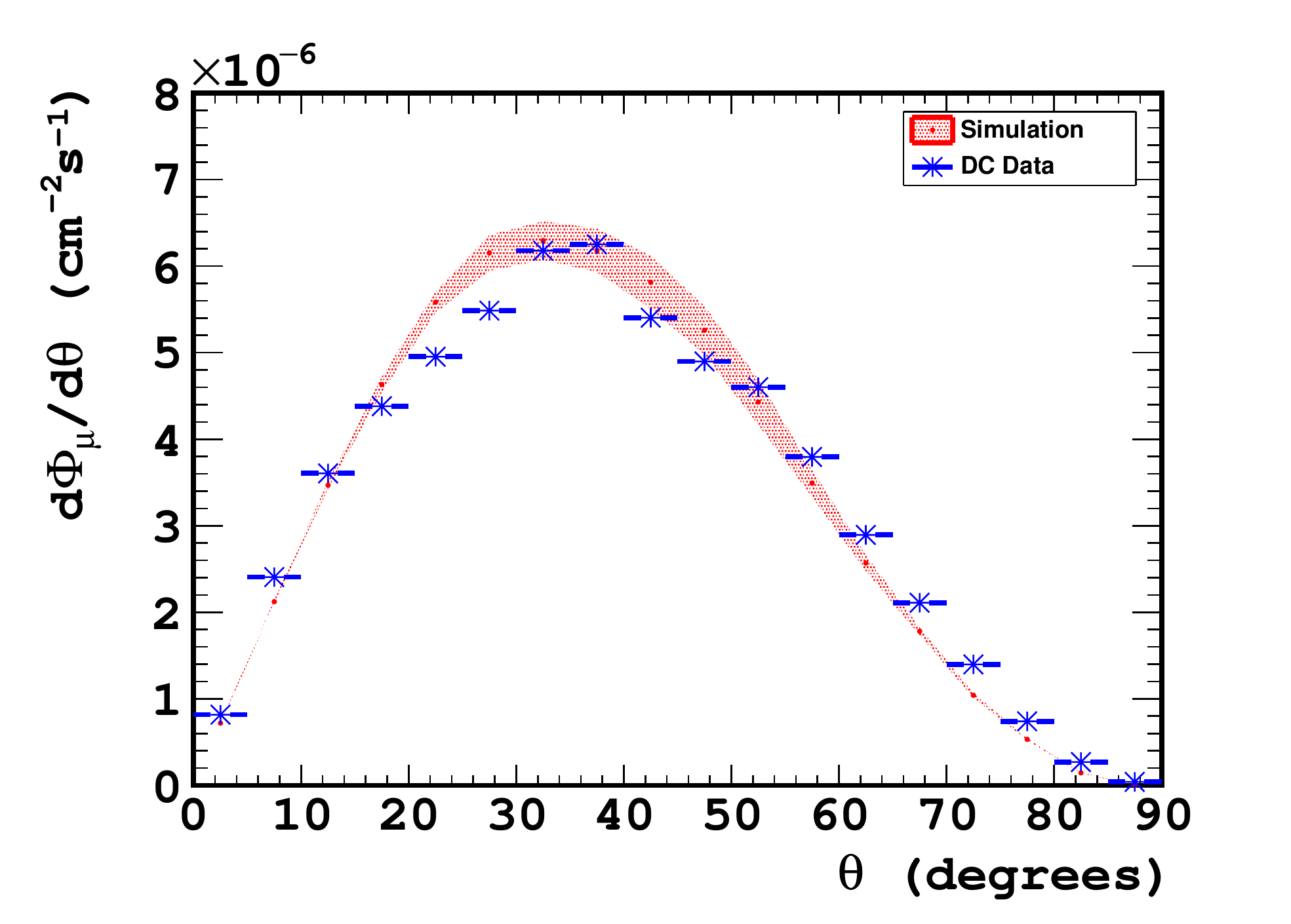}
  \includegraphics[width=0.45\textwidth]{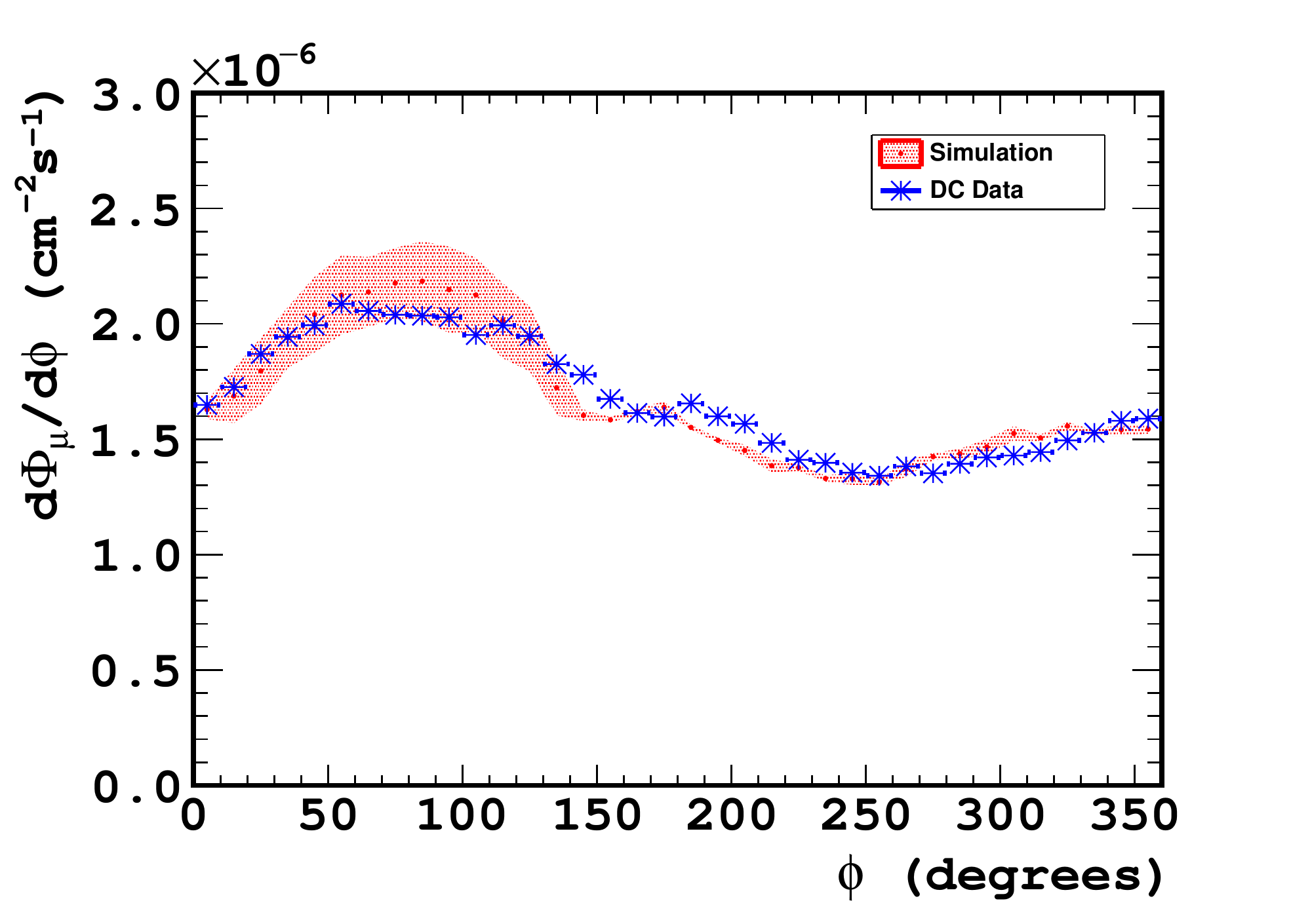}
  \caption{$\theta$ (left) and $\phi$ (right) angular distribution comparisons between the tracks reconstructed from simulations and from experimental data for the Double Chooz near detector.}
  \label{Angular_Comparison_ND}
\end{figure}

\begin{figure}[tbp]
  \centering 
  \includegraphics[width=0.45\textwidth]{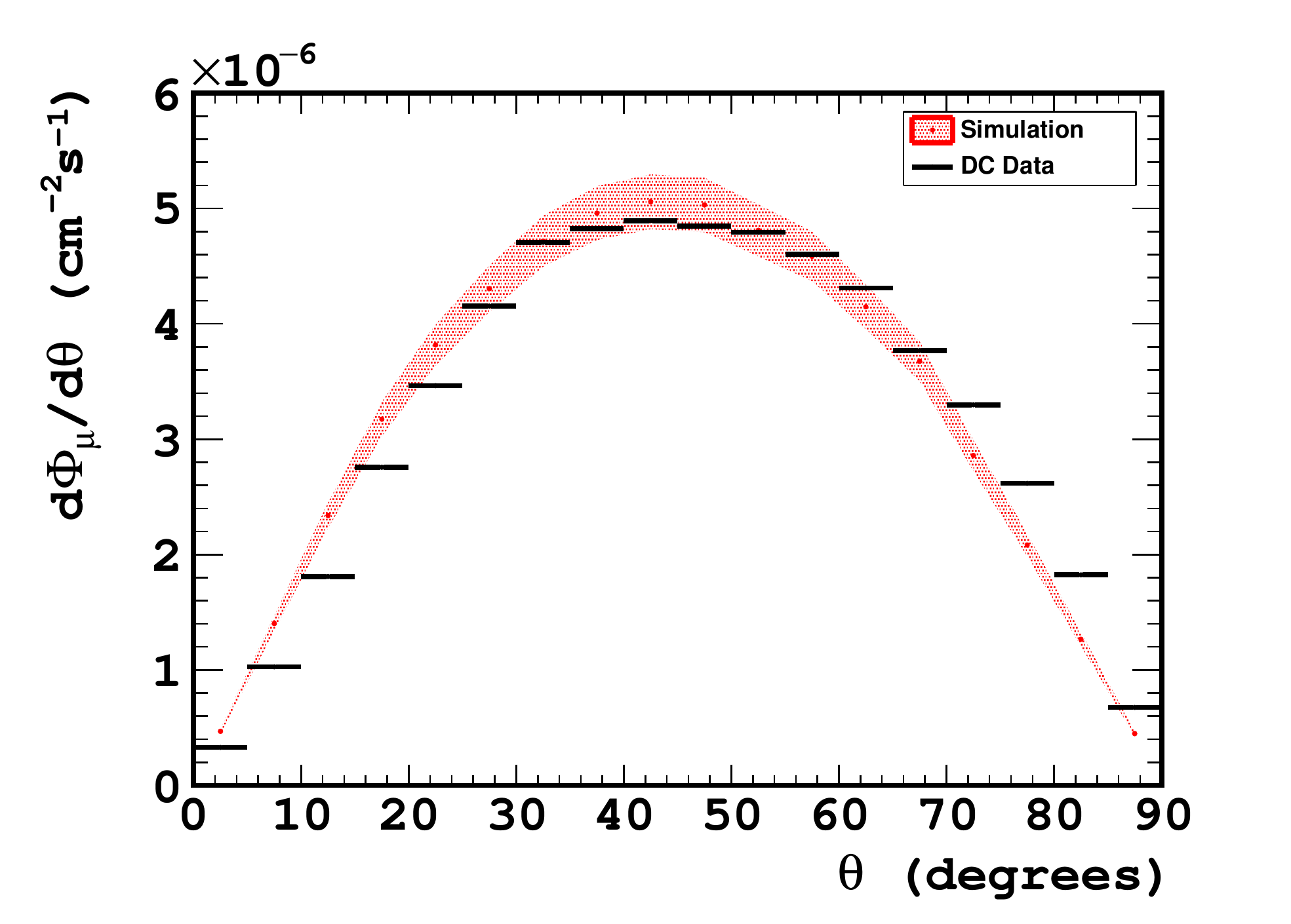}
  \includegraphics[width=0.45\textwidth]{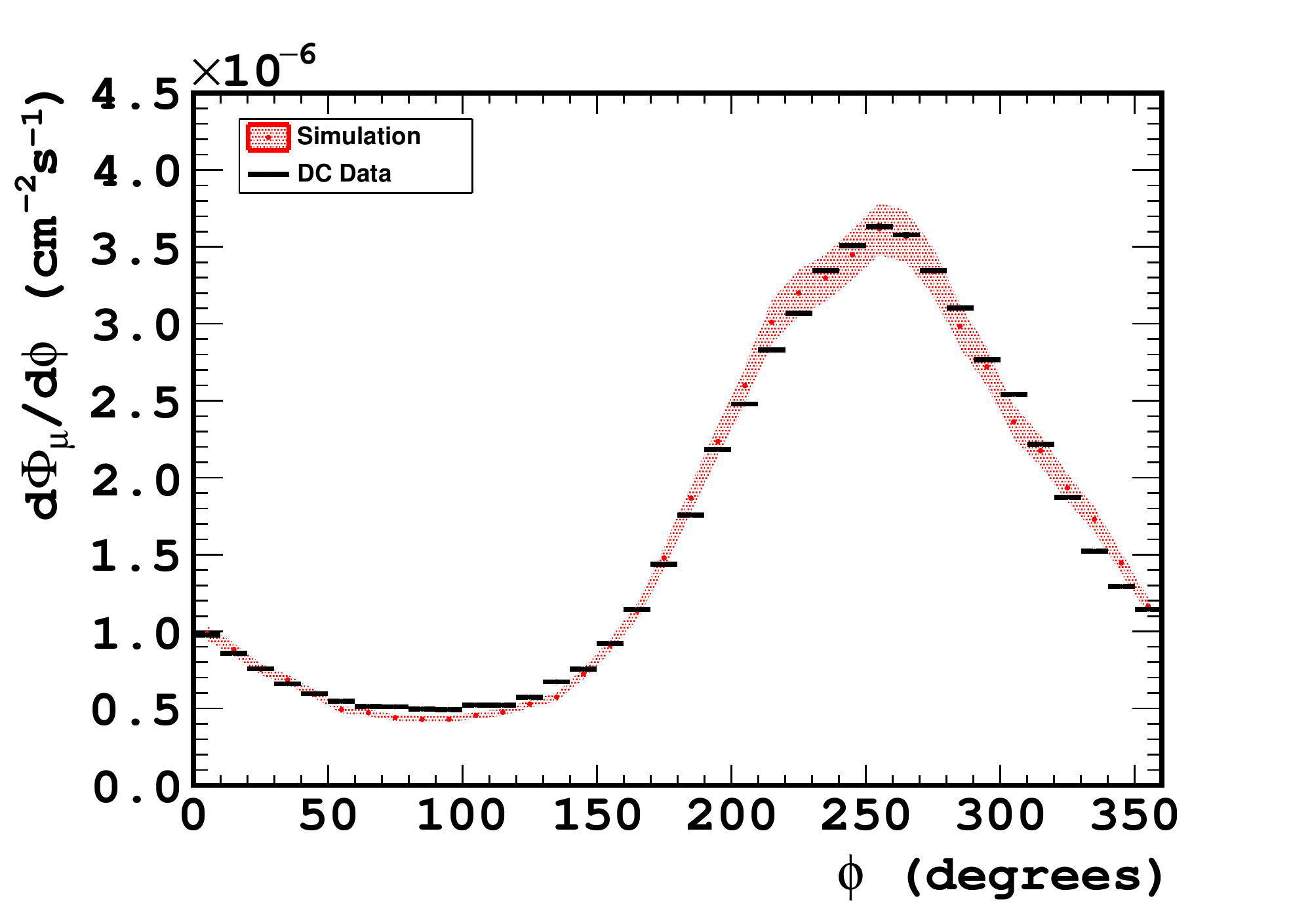}
  \caption{$\theta$ (left) and $\phi$ (right) angular distribution comparisons between the tracks reconstructed from simulations and from experimental data for the Double Chooz far detector.}
  \label{Angular_Comparison_FD}
\end{figure}

Turning to the angular distributions, figure \ref{Muon_output_ThetaPhi} presents the distribution of simulated muons reaching the detectors versus the $\theta$ and $\phi$ angles. As for the reconstructed tracks from data, more populated directions correspond to those where the overburden is thiner. These distributions can be directly compared with those extracted from the experimental data using the muon track reconstruction (figure \ref{Muon_data_output_ThetaPhi}). Figure \ref{Angular_Comparison_ND} shows the zenith and azimuth angular distribution comparison for the near detector, while figure \ref{Angular_Comparison_FD} shows the same comparisons for the far detector case. 

As for the reconstructed muon flux, the muon angular distributions obtained from simulations are also in agreement with those coming from data. For the near detector (figure \ref{Angular_Comparison_ND}), there exist disagreements for some azimuthal directions. The lower precision of the digitized overburden used for simulations translates to larger uncertainties in the results. Moreover, some irregularities can exist in the overburden that cannot be identified with the current precision of the digitized map inducing these effects. However, simulations are in overall agreement with the angular distributions extracted from experimental data. For the case of the far detector (figure \ref{Angular_Comparison_FD}) the overburden has been digitized with higher precision. This leads to smaller uncertainties associated with the simulations than for the near detector. Furthermore, these simulations are compatible with the reconstructed distributions from the experimental data.

\section{Annual modulation of muon flux}
\label{sec:AnnualModulation}

Muons detected by the Double Chooz detectors arise mostly from the decay of charged pions, with a small fraction from the decay of kaons. Mesons are produced by interactions of primary cosmic rays with nuclei of the upper atmosphere and their subsequent decay produces muons. Before reaching the detector, muons lose energy along their path through the atmosphere and through the rock overburden above the laboratories. This implies that muons with the lowest energies are stopped before reaching the detectors. Only higher energy muons can be observed underground, above an energy threshold $E_{\textrm{thr}}$ depending on the depth and profile of the overburden, corresponding to the minimum energy a muon must have at surface to reach the corresponding detector.

For Double Chooz detectors, the energy threshold can be estimated from the simulations performed and described in section \ref{sec:MuonSimulations}. Taking the initial energy spectrum and the correspondig incident angle of muons reaching both detectors, it is possible to compute $\langle E_{\textrm{thr}}\cos\theta\rangle$, being 22.3$\pm$4.8 GeV for the near detector and  46.0$\pm$10.0 GeV for the far one. The use of simulations to obtain these values takes into acount the variations on the energy threshold induced by the ireggularities in the overburden, which leads to obtain different values to those coresponding to the estimated mean depth.

The rate of these detected muons can be related to air density fluctuations in the atmosphere, which in turn affects the fraction of mesons ($\pi$,$K$) decaying to muons. As the average air temperature increases during summer, the average air density is lowered. Due to the less dense medium the mean free path of mesons is longer, increasing the fraction of mesons decaying before having an interaction. Therefore an increase in the temperature of the atmosphere should result in an increase of the measured muon rate. For this reason, a positive correlation between muon rate and averaged atmospheric temperature is expected, leading to seasonal variations with maxima in summer and minima in winter. This effect is known since the early 1950s \cite{RevModPhys.24.133} and has been studied by various experiments located deep underground \cite{2012JCAP...05..015B, Gerda_Ann_Mod, IceCubePaper, 2010PhRvD..81a2001A}.

Other secondary effects, such as a possible correlation with the surface pressure, could also perturb the seasonal variation. This case, also known as barometric effect, is actually a combination of three effects as detailed in \cite{Pham}. Pressure changes also induce density changes in the atmosphere. If the pressure increases, the density does too, leading to additional loss of muons as first effect. Moreover the muon survival probability decreases as second effect. As third effect, with a denser atmosphere more mesons are produced and can eventually decay into muons, so muon production through pions and kaons increases. First two effects reduce the number of muons when the pressure increases, while the third one increases this number. As consequence, the barometric effect implies an overall negative correlation between the muon rate and the surface pressure, defined as the barometer coefficient ($\beta$, expressed in \%/mbar). However, it has been measured \cite{Sagisaka1, Sagisaka2} that this correlation is more relevant for low energy muons and that $\beta$ decreases exponentially for $\langle E_{\textrm{thr}}\cos\theta\rangle$ values above 10 GeV, being of the order of $\sim$10$^{-2}$ \%/mbar for the $\langle E_{\textrm{thr}}\cos\theta\rangle$ values corresponding to Double Chooz detectors. Based on this information, only the correlation between the muon rate variations and the atmosphere effective temperature has been considered in this study.

The effective temperature can be obtained from the temperature profile of the atmosphere. Various measurements are available for this profile. For this study, data provided by the Atmospheric Infrared Sounder (AIRS) instrument of NASA's AQUA satellite have been used. These consist of temperature measurements at a variety of pressure levels (i.e., heights) in the atmosphere, obtained several times a day. From these data, the effective temperature can be estimated from each measurement as \cite{Grashorn2010140}:

\begin{eqnarray}
\label{EffTEquation}
T_{eff}=\frac{\sum_{n=0}^{N-1} \Delta X_n \cdot T(X_n)(W_\pi(X_n)+W_K(X_n))}
{\sum_{n=0}^N \Delta X_n (W_\pi(X_n)+W_K(X_n))},
\end{eqnarray}

\noindent where $\Delta X_n$ is the difference between two adjunct pressure levels, T(X$_{n}$) the corresponding temperature at X$_{n}$ pressure level and $W_{\pi,K}$ the weighting functions of the contributions of pions and kaons to the altitude dependence of the muon production, given by the following expressions:

\begin{eqnarray}\label{eq_WeightningFunctions_1}
W_{\pi,K}(X) & = & \frac{(1-X/\Lambda '_{\pi,K})^2e^{-X/\Lambda_{\pi,K}}A_{\pi,K}^1}{\gamma+(\gamma+1)B_{\pi,K}^1K_{\pi,K}(X)( \langle E_{\textrm{thr}}\cos\theta \rangle /\epsilon_{\pi,K})^2}
\end{eqnarray}

\begin{eqnarray}\label{eq_WeightningFunctions_2}
K_{\pi,K}(X) & = & \frac{(1-X/\Lambda '_{\pi,K})^2}{(1-e^{-X/\Lambda '_{\pi,K}})\Lambda '_{\pi,K}/X}
\end{eqnarray}

\noindent In these equations A$_{\pi,K}^{1}$ includes the amount of inclusive meson production in the forward fragmentation region, masses of mesons and muons, and muon spectral index. B$_{\pi,K}^{1}$ corresponds to the relative atmospheric attenuation of mesons. The critical energies are given by $\epsilon_{\pi,K}$ while the muon spectral index is given by $\gamma$. Finally, $\Lambda_{N}$, $\Lambda_{\pi}$ and $\Lambda_{K}$ represent the attenuation lengths for the cosmic ray primaries, pions, and kaons respectively, with 1/$\Lambda '_{\pi,K}$ $\equiv$ 1/$\Lambda_{N}$ - 1/$\Lambda_{\pi,K}$. The values used for these parameters have been extracted from ref. \cite{2010PhRvD..81a2001A} and summarized in table \ref{tab:InputAlpha}. For this estimation, the nominal value for the kaon to pion ratio $r_{K/\pi}$ = 0.149 $\pm$ 0.060 \cite{gaisser1990cosmic} is used. The $\langle E_{\textrm{thr}}\cos\theta\rangle$ value is 22.3$\pm$4.8 GeV for the near detector and  46.0$\pm$10.0 GeV for the far one as already quoted. This difference on the muon energy threshold for the two detectors induces a difference in the effective temperature smaller than 1\%. The effective temperature per day is obtained by averaging all the measurements performed during that day, taking the standard deviation as the corresponding error.

\begin{table}[tpb]
  \centering
  \begin{tabular}{|l|lc|}\hline
    Parameter & Value & Unit \\
    \hline
    $A_{\pi}^1$ & 1 & -\\
    $A_{K}^1$ & 0.38$\times r_{K/\pi}$ & -\\
    $r_{K/\pi}$ & 0.149$\pm$0.060 & -\\
    $B_{\pi}^1$ & 1.460$\pm$0.007 & - \\
    $B_{K}^1$ & 1.740$\pm$0.028 & - \\
    $\epsilon_\pi$ & 114$\pm$3 & GeV\\
    $\epsilon_K$ & 851$\pm$14 & GeV\\
    $\gamma$ & 1.7$\pm$0.1 & -\\
    $\Lambda_{N}$ & 120 & g/cm$^2$ \\
    $\Lambda_{\pi}$ & 180 & g/cm$^2$ \\
    $\Lambda_{K}$ & 160 & g/cm$^2$ \\
    \hline
  \end{tabular}
  \caption{Input parameters with associated errors for the calculation of the atmospheric effective temperature  $T_{eff}$. Values taken from ref. \cite{2010PhRvD..81a2001A}.}
  \label{tab:InputAlpha}
\end{table}

\begin{figure}[]
\centering
\includegraphics[width=0.8\textwidth]{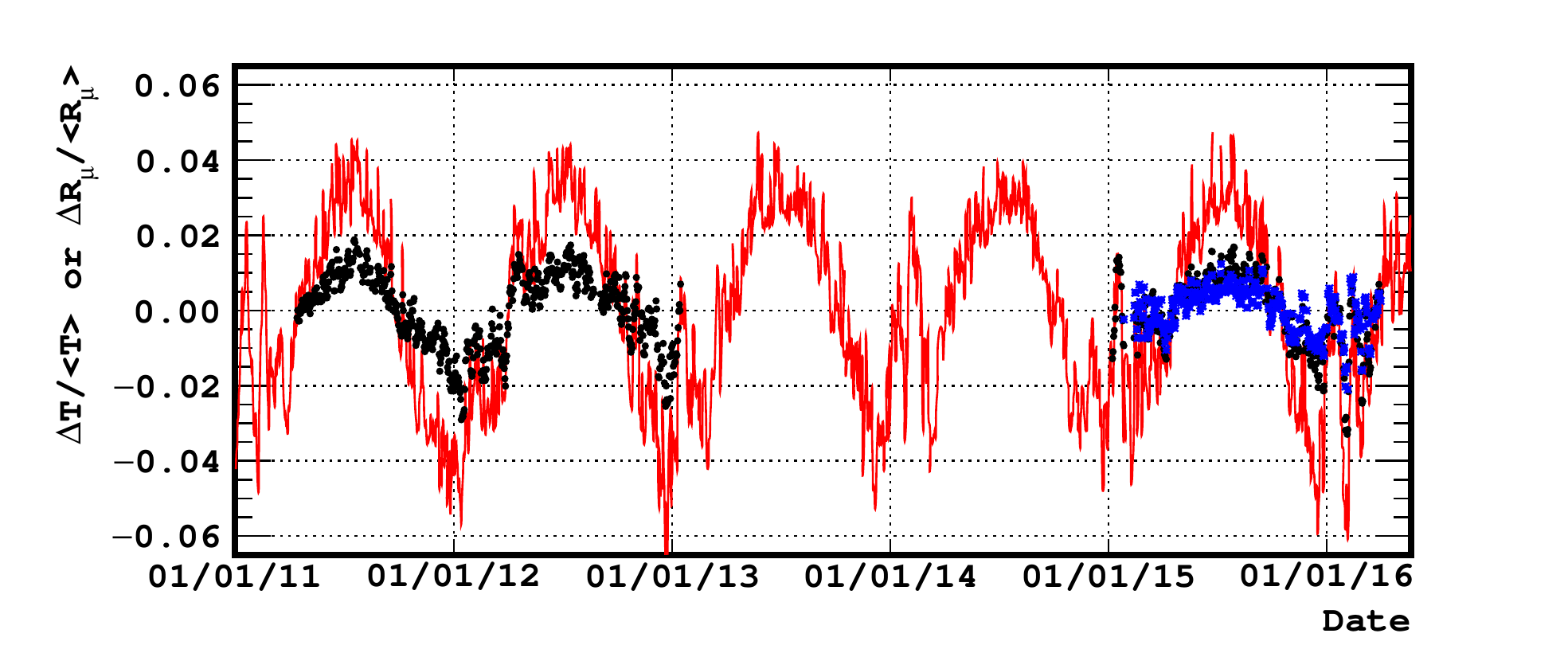}
\caption{Muon rate deviation for near (blue asterisks) and far (black points) detectors together with the effective temperature deviation at Chooz provided by NASA AIRS instrument (red line).}
\label{Mu_T_Correlation}
\end{figure}

The variations of the detected muon rate ($R_{\mu}$) with time can be studied with good statistics, since the far detector has taken data for more than 3 years, and the near detector for over 15 months. For the muon selection, as for the study of the muon angular distribution, the thresholds of the energy deposits in the scintillation volumes of the detectors have been increased in order to obtain a pure sample, setting them to 200 MeV for the ID and 25 MeV for the IV. Since the study is based on rate variations, the associated loss of efficiency does not disturb the obtained results. The variation of the detected muon rate with respect to the mean value for all the data taking periods of Double Chooz are represented in figure \ref{Mu_T_Correlation}\footnote{2013 and 2014 is the period between the end of the data taking with the far detector alone \cite{DChooz} and the beginning of the data taking with the near and far detectors simultaneously. Data from this period of time are not included in this analysis.}. The high rate of muon events allows a significant detection of variations of the order of a few percent. In the same figure the variations of the effective temperature are also showed for the same time period. The seasonal variation of both parameters as well as the correlation between them is obvious. 

By the analysis of the correlation between $\Delta R_{\mu}/\langle R_{\mu} \rangle$ and $\Delta T_{eff} /\langle T_{eff} \rangle$, it is possible to estimate the effective temperature coefficient ($\alpha_{T}$) as \cite{Grashorn2010140}:

\begin{eqnarray}
\label{EffTCoef}
\frac{\Delta R_\mu}{\langle R_\mu\rangle}=\alpha_T\frac{\Delta T_{eff}}{\langle T_{eff}\rangle}.
\end{eqnarray}

To study the correlation, the muon rate and the effective temperature have been averaged over a day during the data taking period. This provides 347 data pairs for the near detector and 897 for the far one. A linear regression between the temperature and rate variations provides the $\alpha_{T}$ value, using the total least squares to take into account errors in both variables. The strength of this linear correlation can be characterized by Pearson's correlation coefficient, which takes values from -1 (perfect negative linear relationship) to +1 (perfect positive linear relationship). 

\begin{figure}
  \centering
    \includegraphics[width=0.45\textwidth]{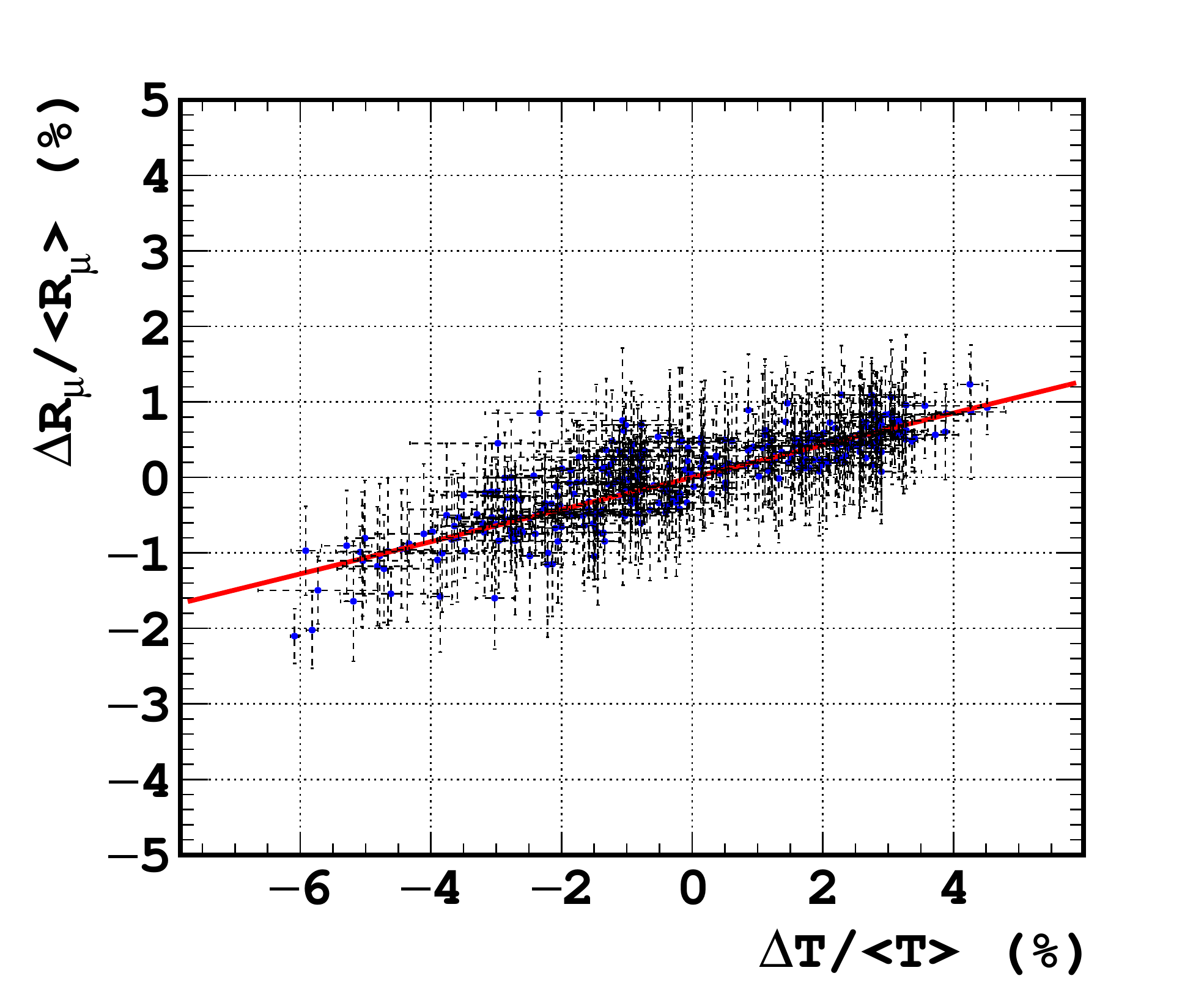}
    \includegraphics[width=0.45\textwidth]{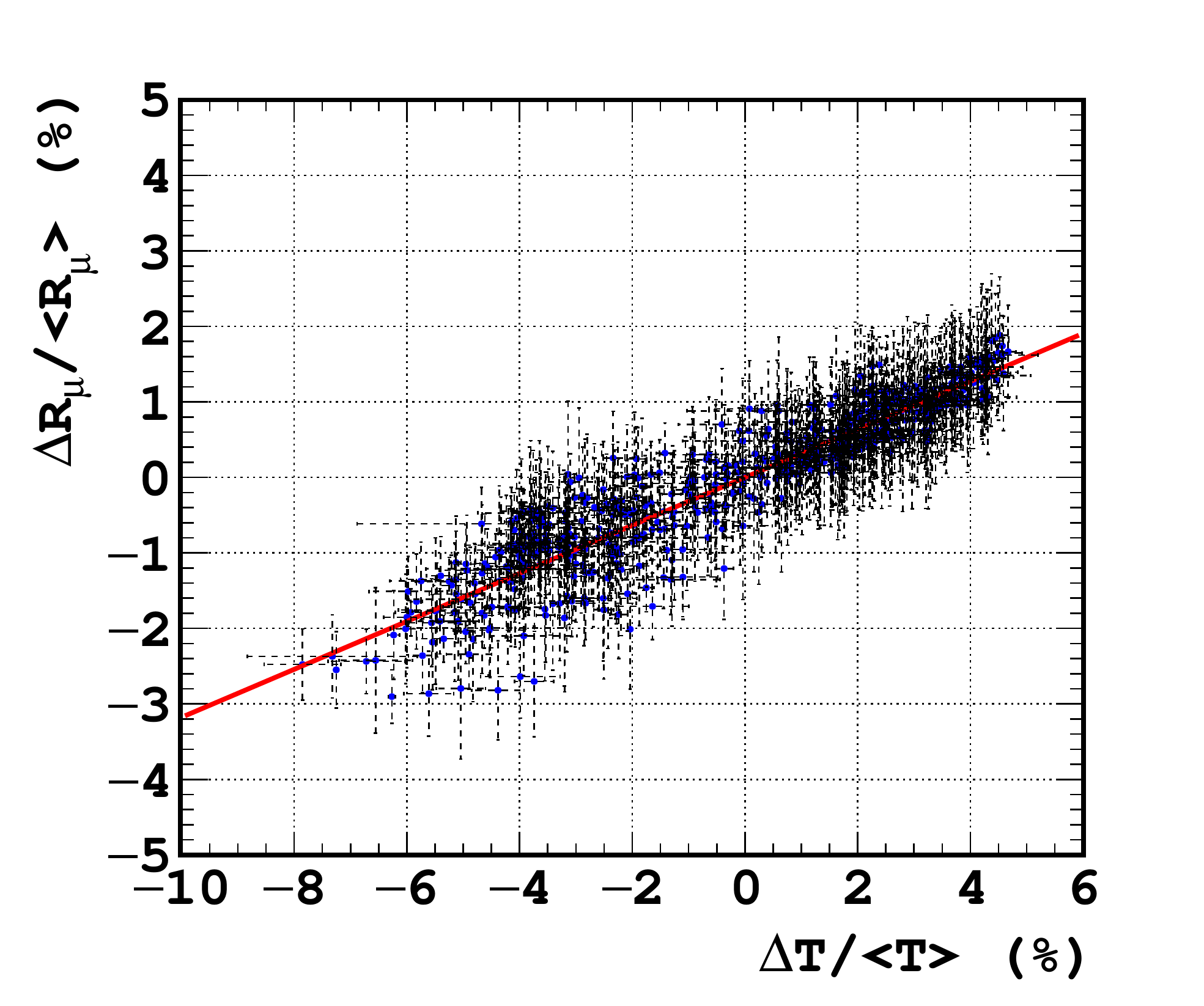}
    \caption{Correlation plots between the muon rate and effective temperature variations for the Double Chooz near (left) and far (right) detectors. Red line corresponds to the linear fit which provides the effective temperature coefficient ($\alpha_{T}$).}
  \label{Correlation_Mu_T}
\end{figure}

Figure \ref{Correlation_Mu_T} shows the correlation between $\Delta R_{\mu}/\langle R_{\mu} \rangle$ and $\Delta T_{eff} /\langle T_{eff} \rangle$ for both Double Chooz detectors, together with their corresponding linear fits, from which the values of $\alpha_{T}$ are extracted. Table \ref{tab:TemperatureCorrelation} summarizes the values of $\alpha_{T}$ and the associated  correlation values estimated for near and far detectors. For $\alpha_{T}$, systematics come from the uncertainties of all the parameters used to estimate the effective temperature as described in eq. \eqref{EffTEquation} and summarized in table \ref{aT_sys}. The main contribution comes from the accuracy of the temperature measurement at different atmosphere levels ($T(X_{n})$). The comparison of these values with those taken by other apparatus impies an accuracy of $\sim$5 \%. The combination of this contribution with all the others provides an overall systematic uncertainty of 5.15 \%.

\begin{table}[]
  \centering
  \begin{tabular}{|l|cc|}
    \hline 
    & $\alpha_{T}$ & Correlation $C(R_\mu, T_{eff})$\\
    \hline
    Near detector & 0.212 $\pm$ 0.013 (stat) $\pm$ 0.011 (sys) & 0.855 \\
    Far detector & 0.355 $\pm$ 0.002 (stat) $\pm$ 0.017 (sys) & 0.923 \\
    \hline 
  \end{tabular}
  \caption{Effective temperature ($\alpha_T$) and correlation ($C(R_\mu, T_{eff})$) coefficients estimated from correlation between the muon rate and effective temperature variations.}
  \label{tab:TemperatureCorrelation}
\end{table}

\begin{table}
\centering
    \begin{tabular}{|l|cc|}
      \hline
      Parameter & Value & $\alpha_T$ systematic contribution\\
      \hline
      $\langle E_{thr}\cos\theta \rangle$ (Near detector) & 22.3$\pm$4.8 GeV & $0.6$\%\\
      $\langle E_{thr}\cos\theta \rangle$ (Far detector) & 46.0$\pm$10.0 GeV & $0.6$\%\\
      $\gamma$ & 1.7$\pm$0.1 & 0.5 \%\\
      $\epsilon_\pi$ & 114$\pm$3 GeV & 0.8 \%\\
      $\epsilon_K$ & 851$\pm$14 GeV & <0.1 \%\\
      $r_{K/\pi}$ &  0.149$\pm$0.060 & <0.1 \%\\
      $B^1_\pi$ & 1.460$\pm$0.007 & 0.2 \%\\
      $B^1_K$ & 1.740$\pm$0.028 & <0.1 \%\\
      $T(X_{n})$ & - & 5.02 \%\\
      \hline
      Total systematic & & 5.15 \%\\
      \hline
    \end{tabular}
    \caption{Summary of the parameters contributing to the systematic error on $\alpha_{T}$. The values of these parameters are quoted together with the contribution to the systematic.}
    \label{aT_sys}
\end{table}

\begin{figure}[t]
\centering
\includegraphics[width=0.6\textwidth]{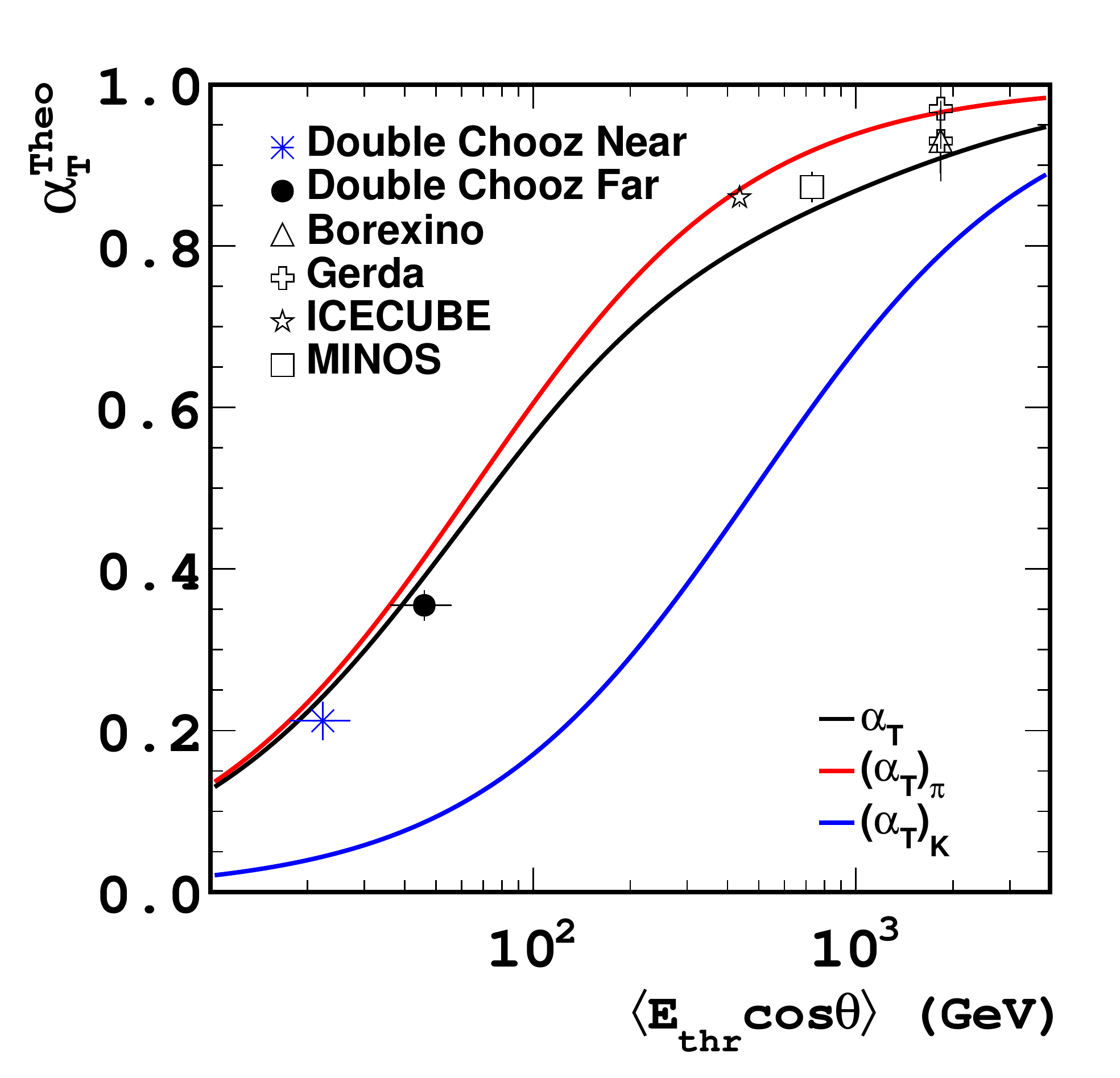}
\caption{Theoretically expected $\alpha_T$ as a function of $\langle E_{\textrm{thr}}\cos\theta \rangle$ in black, where the limit for pions is given in red $(\alpha_T)_\pi$ and the limit for kaons $(\alpha_T)_K$ in blue. $\alpha_{T}$ values estimated for both Double Chooz detectors are also presented together with previous measurements from Borexino \cite{2012JCAP...05..015B}, Gerda \cite{Gerda_Ann_Mod}, ICECUBE \cite{IceCubePaper} and MINOS \cite{2010PhRvD..81a2001A}.}
\label{AlphaTth}
\end{figure}

Correlation values confirm the positive linear relationship between the muon rate and temperature variations. Regarding $\alpha_{T}$, the value for the near detector is smaller than for the far one as expected due to the lower overburden which allows lower energy muons to reach the detector. Both values can be compared with the theoretical value of this coefficient $\alpha^{\textrm{Theo}}_{T}$, which can be estimated as:

\begin{eqnarray}
\label{alphaT_Teo1}
\alpha_{T}^{\textrm{Theo}}=\frac{1}{D_{\pi}}\frac{1/\epsilon_{K}+A_{K}^{1}(D_{\pi} /D_{K})^2/\epsilon_{\pi}}{1/\epsilon_{K}+A_{K}^{1}(D_{\pi} /D_{K})/\epsilon_{\pi}}
\end{eqnarray}

\begin{eqnarray}
\label{alphaT_Teo2}
D_{K,\pi}=\frac{\gamma}{\gamma+1}\frac{\epsilon_{K,\pi}}{1.1 \langle E_{thr}\cos\theta \rangle}+1
\end{eqnarray}

\noindent where the parameters are those used in eqs. \eqref{eq_WeightningFunctions_1} and \eqref{eq_WeightningFunctions_2}. Figure \ref{AlphaTth} shows this theoretical value together with the limits from kaons and pions (i.e., assuming that only the corresponding particle contributes to the seasonal effect) considering $r_{K/\pi}$ = 0.149 $\pm$ 0.060 \cite{gaisser1990cosmic}. The Double Chooz values for both detectors are also represented together with measurements of other experiments. The extracted values for the Double Chooz detectors provide an experimental estimation of the effective temperature coefficient at low energies and are compatible with the theoretical value taking into account the associated uncertainties, validating the theoretical model, and by extension the $r_{K/\pi}$ nominal value,  in the low $\langle E_{\textrm{thr}}\cos\theta \rangle$ range. These results complement those previously obtained by other underground experiments for higher $\langle E_{\textrm{thr}}\cos\theta \rangle$ values as showed in figure \ref{AlphaTth}. 

\section{Conclusions}
\label{sec:Conclusions}

Characterization of muons reaching the Double Chooz detectors has been performed using the large quantity of data available from the experiment, which comprises $\sim$151 days for the near detector and $\sim$673 days for the far one. The combination of the muon selection, based on the energy released by the muons in the various volumes of the detectors, and the muon track reconstruction performed by a dedicated algorithm, allows the estimation of the mean muon flux of (3.64 $\pm$ 0.04) $\times$ 10$^{-4}$ cm$^{-2}$s$^{-1}$ and (7.00 $\pm$ 0.05) $\times$ 10$^{-5}$ cm$^{-2}$s$^{-1}$ for the near and far detectors respectively. Moreover, the angular distributions of reconstructed muon tracks show correlations between the detected muon rate and the overburden profile, identifying privileged directions where the overburden is shallower. This effect is more evident for the far detector, with its more irregular overburden. 

In parallel, a simulation for both detectors has been performed using MUSIC. From these simulations it has been made possible to reconstruct the muon flux reaching the detector as well as the angular distributions. For both detectors, the reconstructed flux and angular distributions are in agreement with the experimental data. Some discrepancies in the near detector case are directly related to the lower precision of the overburden model used, together with the fact that the detector is traversed by lower energy muons for which the models are more uncertain.

Finally, taking advantage of the high statistics accumulated for both detectors, the seasonal variation of the detected muon rate and its correlation with the atmospheric effective temperature has been studied. The effective temperature coefficient has been estimated for both detectors giving 0.212$\pm$0.024 for the near detector and 0.355$\pm$0.019 for the far one. Both values are compatible with theoretical expectations based on the nominal value for the kaon to pion ratio ($r_{K/\pi}$), and taking into account the associated uncertainties. These results, represent one of the first measurements of this parameter in shallow depth installations, and therefore for a lower muon energy than other underground experiments. 

\acknowledgments

We thank the French electricity company EDF; the European fund FEDER;
the R\'egion de Grand Est; the D\'epartement des Ardennes;
and the Communaut\'e de Communes Ardenne Rives de Meuse.
We acknowledge the support of the CEA, CNRS/IN2P3, the computer centre CCIN2P3, and UnivEarthS Labex program of Sorbonne Paris Cit\'e in France (ANR-10-LABX-0023 and ANR-11-IDEX-0005-02);
the Ministry of Education, Culture, Sports, Science and Technology of Japan (MEXT) and the Japan Society for the Promotion of Science (JSPS);
the Department of Energy and the National Science Foundation of the United States;
U.S. Department of Energy Award DE-NA0000979 through the Nuclear Science and Security Consortium;
the Ministerio de Econom\'ia y Competitividad (MINECO) of Spain;
the Max Planck Gesellschaft, and the Deutsche Forschungsgemeinschaft DFG, the Transregional Collaborative Research Center TR27, the excellence cluster ``Origin and Structure of the Universe'', and the Maier-Leibnitz-Laboratorium Garching in Germany;
the Russian Academy of Science, the Kurchatov Institute and RFBR (the Russian Foundation for Basic Research);
the Brazilian Ministry of Science, Technology and Innovation (MCTI), the Financiadora de Estudos e Projetos (FINEP), the Conselho Nacional de Desenvolvimento Cient\'ifico e Tecnol\'ogico (CNPq), the S\~ao Paulo Research Foundation (FAPESP), and the Brazilian Network for High Energy Physics (RENAFAE) in Brazil.


\begin{thebibliography}{99}

\bibitem{DChooz} Y. Abe et al. (Double Chooz Collaboration), \emph{Improved measurements of the neutrino mixing angle $\theta_{13}$  with the Double Chooz detector},  \emph{JHEP} \textbf{10} (2014) 086.

\bibitem{DChooz_LN} Y. Abe et al. (Double Chooz Collaboration), \emph{Characterization of the spontaneous light emission of the PMTs used in the Double Chooz experiment},  \emph{JINST} \textbf{11} (2016) P08001.

\bibitem{FIDO} Y. Abe et al. (Double Chooz Collaboration), \emph{Precision muon reconstruction in Double Chooz},  \emph{Nucl. Inst. and Meth. A} \textbf{764} (2014) 330.

\bibitem{MUSIC} P. Antonioli et al., \emph{A three-dimensional code for muon propagation through the rock: MUSIC}, \emph{Astroparticle Physics} \textbf{7} (1997) 357.

\bibitem{3DField} V. Galouchko, \url{http://field.hypermart.net}

\bibitem{IGN} \url{http://www.ign.fr/} and \url{http://professionnels.ign.fr/rgealti}

\bibitem{gaisser1990cosmic} T.K. Gaisser, \emph{Cosmic rays and particle physics}, 
Cambridge University Press (1990).

\bibitem{Tang_PhysRevD_2006} A. Tang et al., \emph{Muon simulations for Super-Kamiokande, KamLAND, and CHOOZ}, \emph{Physical Review D} \textbf{74} (2006) 053007.

\bibitem{RevModPhys.24.133} P.H. Barret, \emph{Interpretation of Cosmic-Ray Measurements Far Underground}, \emph{Rev. Mod. Phys.} \textbf{24} (1952) 133.

\bibitem{2012JCAP...05..015B} G. Bellini et al., \emph{Cosmic-muon flux and annual modulation in Borexino at 3800 m water-equivalent depth}, \emph{JCAP} \textbf{2012 - 05} (2012).

\bibitem{Gerda_Ann_Mod} M. Agostini et al., \emph{Flux modulations seen by the muon veto of the Gerda experiment}, \emph{Astroparticle Physics} \textbf{84} (2016) 29.

\bibitem{IceCubePaper} S. Tilav et al., \emph{Atmospheric Variations as observed by IceCube}, 
arXiv: 1001.0776.

\bibitem{2010PhRvD..81a2001A} P. Adamson et al., \emph{Observation of muon intensity variations by season with the {MINOS} far detector}, \emph{Phys. Rev. D} \textbf{81} (2010) 012001.

\bibitem{Pham} P.N. Diep et al., \emph{Dependence of the cosmic muon flux on atmospheric pressure and temperature}, \emph{Communications in Physics} \textbf{14} (2004) 57.

\bibitem{Sagisaka1} S. Sagisaka et al., \emph{Atmospheric effect on muon intensity for observations at various atmospheric and underground depths}, \emph{16th ICRC} \textbf{4} (1979) 235.

\bibitem{Sagisaka2} S. Sagisaka et al., \emph{Atmospheric effects on cosmic-ray muon intensities at deep underground depths}, \emph{Il Nuovo Cimento C} \textbf{9} (1986) 809.

\bibitem{Grashorn2010140} E.W. Grashorn, \emph{The atmospheric charged kaon/pion ratio using seasonal variation methods}, \emph{Astroparticle Physics} \textbf{33} (2010) 140.

\end{thebibliography}
\end{document}